\documentclass[prd,preprint,superscriptaddress,preprintnumbers,eqsecnum,showpacs,nofootinbib,nobibnotes]{revtex4}
\usepackage{amsfonts,amsmath,amssymb,bm,natbib}
\usepackage{graphicx} 
\newcommand{\be}{\begin{equation}}
\newcommand{\bea}{\begin{eqnarray}}
\newcommand{\ba}{\begin{align}}
\newcommand{\ee}{\end{equation}}
\newcommand{\eea}{\end{eqnarray}}
\newcommand{\ea}{\end{align}}
\def\s#1{{\scriptscriptstyle #1}}

\def\1eq#1{Eq.~(\ref{#1})}

\def\2eqs#1#2{Eqs.~(\ref{#1}) and~(\ref{#2})}
\def\3eqs#1#2#3{Eqs.~(\ref{#1}),~(\ref{#2}) and~(\ref{#3})}
\def\noeq#1{(\ref{#1})}

\def\G{\Gamma} 
\def\deltag{\delta h}

\def\Gz{\Gamma^{(0)}}
\def\Gzt{\widetilde{\Gamma}^{(0)}}
\def\Gt{\widetilde{\Gamma}}

\def\d{\mathrm d}

\def\spr{\!\cdot\!}

\begin{document}

\title{The Cosmological Slavnov-Taylor Identity\\from BRST Symmetry in Single-Field Inflation}
\date{November 15, 2015}

\author{D. Binosi}
\email{binosi@ectstar.eu}
\affiliation{European Centre for Theoretical Studies in Nuclear
  Physics and Related Areas (ECT*) and Fondazione Bruno Kessler, \\ Villa Tambosi, Strada delle
  Tabarelle 286, 
 I-38123 Villazzano (TN)  Italy}
 
\author{A. Quadri}
\email{andrea.quadri@mi.infn.it}
\affiliation{Dip. di Fisica, Universit\`a degli Studi di Milano, via Celoria 16, I-20133 Milano, Italy\\
and INFN, Sezione di Milano, via Celoria 16, I-20133 Milano, Italy}

\begin{abstract}
\noindent
The cosmological Slavnov-Taylor (ST) identity of the Einstein-Hilbert action coupled to a single inflaton field is obtained from the Becchi-Rouet-Stora-Tyutin (BRST) symmetry associated with diffeomorphism invariance in the Arnowitt-Deser-Misner (ADM) formalism. The consistency conditions between the correlators of the scalar and tensor modes in the squeezed limit are then derived from the ST identity, together with the softly broken conformal symmetry. Maldacena's original relations connecting the  2- and 3-point correlators at horizon crossing are recovered, as well as the next-to-leading corrections, controlled by the special conformal transformations.
\end{abstract}


\maketitle

\section{Introduction}

Non-gaussianities of the CMB spectrum in single field inflationary models are described by correlators involving gravitational scalar and tensor modes that fulfil a set of consistency conditions first derived in a seminal paper by Maldacena~\cite{Maldacena:2002vr}. The latter hold in a particular limit (the so-called squeezed limit), where one of the momenta is much smaller than the others.

Recently, there has been a renewed interest in providing a rigorous mathematical derivation of these (and associated) relations in terms of cosmological Slavnov-Taylor (ST) identities, {\it i.e.} functional identities for the connected inflationary generating functional. So far the approach has been the one of translating at the path-integral level diffeomorphism invariance or (softly broken) conformal symmetry~\cite{Collins:2014fwa,Armendariz-Picon:2014xda,Goldberger:2013rsa,Berezhiani:2013ewa}, thus obtaining the aforementioned identities through formal manipulations of the 
(gauge-fixed) path-integral for each symmetry.

However, from the algebraic point of view the situation does not seem to be very satisfactory. The gauge invariance of the Einstein-Hilbert action coupled to the inflaton field is broken by the gauge-fixing and it should be therefore replaced by the corresponding Becchi-Rouet-Stora-Tyutin (BRST) symmetry~\cite{Becchi:1974md,Becchi:1975nq,Becchi:1996yh,Tyutin:1975qk}. The latter, in turn, should yield a single ST identity~\cite{Slavnov:1972fg,Taylor:1971ff} which is expected to encode all the relevant consistency relations between the correlators of the tensor and scalar gravitational modes.

In this paper we prove that this is indeed the case. Starting from the Arnowitt-Deser-Misner (ADM) formulation of the gravitational action~\cite{Arnowitt:1962hi,Arnowitt:1960es}, we derive the BRST symmetry associated with diffeomorphism invariance. A careful discussion of the lapse and shift vector (that are not dynamical and can be eliminated through their equations of motion) is worked out. Next, as is customary in the gauge-fixing procedure {\em \`a la} BRST, we identify the relevant set of external sources (antifields) required to define the composite operators entering into the BRST transformation of the fields~\cite{Gomis:1994he}. This allows to finally derive the (unique) ST identity associated with BRST symmetry, which, under functional differentiations with respect to appropriate field and antifield combinations, give rise to relations between the one-particle irreducible (1-PI) Green functions of the theory. One of the advantages of the resulting functional formulation is that it can be generalized to study quantum corrections to the tree-level amplitudes, although in the present paper we will not attempt to go beyond the classical approximation.

The direct application of this approach is however hampered by the fact that in cosmology one is usually interested in correlation functions evaluated at a fixed time, usually coinciding with horizon crossing. Such correlators are computed by using the so-called ``in-in'' formalism~\cite{Weinberg:2005vy}. We therefore introduce the corresponding generating functional $W(t)$  and derive the ST identity it fulfils, as a consequence of the BRST invariance of the classical gauge-fixed action. We then motivate the need to introduce the 1-PI generating functional at fixed time $\Gzt(t)$, as a necessary tool in order to recover the consistency conditions in the squeezed limit for the soft momentum~$\vec{q}$. In this respect, one should notice that the reconstruction of the 1-PI vertex functional from $W(t)$ is carried out once one has integrated out the lapse and shift vector. This in turn introduces some sources of non-locality, despite the fact that the Einstein-Hilbert action in the ADM formalism is local. However, for an attractor background~\cite{Cheung:2007sv} under the adiabaticity assumption, guaranteeing that the growing modes solution of the classical equations of motion are constant, one can see that such non-localities disappear~\cite{Berezhiani:2013ewa}.

When these conditions are met, suitable analyticity assumptions of the 1-PI Green's functions can be made, which allows, once one factors out the two point correlator of the soft momentum $\vec{q}$, to constrain the remaining contribution to scalar and/or graviton correlators ``in-in'' amplitudes through the ST identity. Indeed, the constant and linear term in $\vec{q}$ of the expansion of the amplitudes for $\vec{q} \rightarrow 0$ can be traced back to the invariance under dilatations~\cite{Maldacena:2002vr} and special conformal transformations~\cite{Creminelli:2012ed} respectively. In our approach they can be both derived from a single ST identity. In addition, the particular form of the BRST symmetry ensures that there are no graviton contributions for purely scalar functions at the leading and next-to-leading order in $\vec{q}$.

Thus, the BRST formalism introduced provides a unified approach to obtain the expansion of the relevant Green's functions as $\vec{q}$ goes to zero.

The paper is organized as follows. In Sect.~\ref{sec:ADM.action} we review the ADM formulation of gravity coupled to a single scalar field and set up our notations. In Sect.~\ref{sec.brst.gf} we derive the BRST symmetry of the theory, introduce the gauge-fixing as well as the ghost sector. The ensuing ST identity is derived in Sect.~\ref{sec.ST}. In Sect.~\ref{sec.maldacena} the spacetime symmetries are formulated within the BRST approach and it is explained how dilatations and special conformal transformations are recovered from BRST symmetry. This paves the way to the derivation in Sect.~\ref{sec:cosmological.sti} of the cosmological ST identity fulfilled by the generating functional of correlators at horizon crossing in the ``in-in'' formalism. The results of~Refs.\cite{Maldacena:2002vr} and \cite{Creminelli:2012ed} on the consistency conditions at the leading and next-to-leading order in the squeezed limit for 2- and 3-point scalar and graviton correlators are also reproduced. Finally, our conclusions are presented in Sect.~\ref{sec.mcc}. The paper ends with some Appendices, devoted to clarify some technical aspects of the derivations, and in particular: the diagonalization of the quadratic ADM action (Appendix~\ref{app.quadratic.part.adm}); the derivation of the metric propagator (Appendix~\ref{app.metric.prop}); the functional identities for the effective action and the connected generating functional (Appendix~\ref{app.funct}); the consistency conditions in functional form (Appendix~\ref{app.gauge.indip.mast}).

\section{ADM Action}\label{sec:ADM.action}

The Einstein-Hilbert action for gravity coupled to a single scalar field reads
\begin{align}
S = \frac{1}{2} \int \! \sqrt{g}\, \left[R - (\nabla \phi)^2 - 2 V(\phi) \right],
\label{action}
\end{align}
where, using the same notations as in~\cite{Maldacena:2002vr}, we have set $M^{-2}_{\rm Planck} =  8 \pi G_N = 1$.

Passing to the ADM formulation, one introduces the lapse function ${\cal N}$ and shift vector ${\cal N}^i$ which are implicitly defined through the metric decomposition
\begin{align}
	\d s^2&=-{\cal N}^2\d t^2+h_{ij}\left(\d x^i+{\cal N}^i\d t\right)\left(\d x^j+{\cal N}^j\d t\right).
	\label{adm.metric.decomp}
\end{align}
The action (\ref{action}) can be then written as the sum of three terms, $S=S_1+S_2+S_3$,
with
\begin{align}
	S_1 &= \frac12\int\!{\mathrm d}t\!\int\! {\mathrm d}^3 \vec{x} \, \sqrt{h}\, {\cal N}\, ^{(3)}\! R,\nonumber \\
	S_2 &= \frac{1}{2} \int\!dt\int\! d^3 \vec{x} \, \sqrt{h}\, {\cal N}^{-1} (E_{ij} E^{ij} - E^2);\qquad
	E_{ij} = \frac{1}{2} ( \dot\deltag_{ij} - \nabla_i\, {\cal N}_j - \nabla_j\, {\cal N}_i );\quad E = E^i_i,\nonumber \\
	S_3 &= \frac{1}{2} \int\!{\mathrm d}t\!\int\! {\mathrm d}^3 \vec{x} \, \sqrt{h}\, \left[{\cal N}^{-1} (\dot{\phi} - {\cal N}^{\,i} \partial_i \phi)^2-{\cal N} h^{ij} \partial_i  \phi \partial_j \phi-2{\cal N} V(\phi)\right].
	\label{ADMSi}
\end{align}

The homogenous solution of this action is obtained for ${\cal N}=1$, ${\cal N}^i=0$, $h_{ij}=\widehat{h}_{ij}=e^{2\rho}\delta_{ij}$, and corresponds to a de Sitter (dS) spacetime. In addition, $\rho=\rho(t)$ and $\phi=\varphi(t)$ depends only on time, and satisfy the equations of motion
\begin{align}
	3\dot{\rho}^2&=\frac12\dot{\varphi}^2+V(\varphi);&
	\ddot{\rho}&=-\frac12\dot{\varphi}^2;&
	\ddot{\varphi}+3\dot\rho\dot{\varphi}+V'(\varphi)=0.
	\label{eom1}
\end{align}
The Hubble parameter is then given by $H \equiv \dot{\rho}$, whereas
the slow roll parameters can be defined according to
\begin{align}
	\epsilon&=\frac12\left(\frac{V'}{V}\right)^2\sim\frac12\frac{\dot\varphi^2}{\dot\rho^2};&
	\eta &=\frac{V''}{V}\sim-\frac{\ddot\varphi}{\dot\rho\dot\varphi}+\frac12\frac{\dot\varphi^2}{\dot\rho^2}.
	\label{slowroll}
\end{align}
The approximate relations above hold when the slow roll parameters are small, which will in turn lead to a period of accelerated expansion.

In the ensuing analysis, we consider quantum fluctuations around the dS background solution. We will choose a general gauge, defined through
\begin{align}
	h_{ij} &= \widehat{h}_{ij} +e^{2\rho} \deltag_{ij};& \phi&=\varphi+\delta\phi;& {\cal N}=1+\delta{\cal N},
	\label{gengauge}
\end{align}
and expand the ADM action up to quadratic orders in the various fields.  The straightforward but somehow lengthy computations are summarized in Appendix~\ref{app.quadratic.part.adm}, where it is also shown that, upon substitution of the solutions of the equations of motion for the lapse function and the shift vector, one gets back, after choosing the appropriate gauge, the same results reported in~\cite{Maldacena:2002vr}.
 
In the gauge-fixing {\em \`a la} BRST we are going to carry out in the next Section,
one needs to keep all degrees of freedom, including the non-dynamical ones. Propagators are obtained as usual
by inversion of the 2-point sector of the classical action after gauge-fixing. For that purpose it is convenient
to diagonalize the quadratic part of the action (see Appendix~\ref{app.quadratic.part.adm}). Elimination
of the non-dynamical lapse and shift vector in the effective action is equivalent to consider diagrams that are one-particle reducible with respect to such fields.

\section{BRST symmetry and gauge-fixing}\label{sec.brst.gf}

In a gravitational theory gauge transformations correspond to (minus) the Lie derivative with respect to the vector field of gauge parameters $\xi$, that is one has
\begin{align}
	\delta_\xi\Phi=-\underset{\xi}{\cal L}\Phi;\qquad \Phi={\cal N},{\cal N}_i, h_{ij},\phi.
\end{align}
Since ${\cal N}=(-g^{00})^{1/2}$ and ${\cal N}^i=-g^{0i}/g^{00}$, the individual transformations of all the fields $\Phi$ can be obtained starting from the gauge transformation of the four-dimensional metric, which reads
\begin{align}
	\delta_\xi g_{\mu\nu}=-\xi^\rho\nabla_\rho g_{\mu\nu}-g_{\mu\rho}\nabla_\nu\xi^\rho+g_{\nu\rho}\nabla_\mu\xi^\rho.
	\label{diffeo-g}
\end{align}
One then gets
\begin{align}
	\delta_\xi{\cal N}&=-\nabla_0(\xi^0{\cal N})+{\cal N}{\cal N}^k\nabla_k\xi^0-\xi^k\nabla_k\,{\cal N},\nonumber \\
	\delta_\xi{\cal N}^i&={\cal N}^2h^{ij}\nabla_j\xi^0-\nabla_0\xi^i+{\cal N}^j\nabla_j\xi^i-\xi^0\nabla_0{\cal N}^i-{\cal N}^i\nabla_0\xi^0+{\cal N}^i{\cal N}^j\nabla_j\xi^0,\nonumber \\
	\delta_\xi h_{ij}&=-\xi^0\nabla_0h_{ij}-\xi^m\nabla_m h_{ij}-h_{im}{\cal N}^m\nabla_j\xi^0-h_{jm}{\cal N}^m\nabla_i\xi^0-h_{im}\nabla_j\xi^m-h_{jm}\nabla_i\xi^m,\nonumber \\
	\delta_\xi\phi&=-\xi^\mu\partial_\mu\phi.
	\label{brst.full}
\end{align}
In fact, it should be noticed that, as the metric is torsion-free, the covariant derivatives can be replaced by conventional ones, so that~\noeq{diffeo-g} becomes
\begin{align}
	\delta_\xi g_{\mu\nu}=-\xi^\rho\partial_\rho g_{\mu\nu}-g_{\mu\rho}\partial_\nu \xi^\rho-g_{\nu\rho}\partial_\mu \xi^\rho,
	\label{diffeo-g-simple}
\end{align} 
with similar simplified expressions obtained (through $\nabla\to\partial$) holding for the individual gauge transformations~\noeq{brst.full}.

The BRST transformations are then generated by replacing in~\2eqs{brst.full}{diffeo-g-simple} the gauge parameter vector field with the ghost field, $\xi\to c$, and correspondingly $\delta_\xi\to s$ (with $s$ denoting the BRST differential). The action of the BRST differential on the ghosts is finally fixed by demanding the fulfilment of the nilpotency condition; indeed, applying $s$ on~\1eq{diffeo-g-simple} and requiring that $s^2=0$ we get 
\begin{align}
s c^\mu = -c^\nu\partial_\nu c^\mu.
\label{brst.ghost}
\end{align}

The inversion of the quadratic part of the action~\noeq{ADMSi} requires to fix a gauge. In our BRST approach this is achieved by supplying a gauge fixing condition ${\cal F}_\mu$ and coupling it with a corresponding set of Nakanishi-Lautrup multipliers~\cite{Nakanishi:1966zz,Lautrup:1967zz}.

While the procedure is completely general, in order to fix the ideas we apply it after  choosing a comoving gauge for the inflaton field
\begin{align}
{\cal F}_0 \equiv \delta \phi = 0,
\label{comov.gauge.infl}
\end{align}
as well as a transverse gauge-fixing condition which is reminiscent of the Landau gauge often employed in Yang-Mills theories, {\it i.e.},%
\begin{align}
	{\cal F}_j \equiv \partial^i \deltag_{ij}-\frac13\partial_j \deltag^i_i=0.
	\label{como_trans}
\end{align}
Notice that at the classical level the metric $\delta h_{ij}$ can be written without loss of generality as
\begin{align}
	\delta h_{ij}&=2\psi\delta_{ij}+2\partial_i\partial_j E+\partial_iF_j+\partial_jF_i+\gamma_{ij};& \partial^i F_i&=0;& \partial^i\gamma_{ij}&=0;\ \gamma_i^i=0.
\label{metric.decomp}
\end{align}
The transverse condition Eq.~(\ref{como_trans}) then implies, disregarding zero modes of the three-dimensional Laplacian (as in comoving gauge), $F_i=E=0$. 

The gauge fixing conditions~\noeq{comov.gauge.infl}  and \noeq{como_trans} are then implemented by coupling them to the auxiliary, non-dynamical fields $b^0$ and $b^i$ respectively (representing the aforementioned Nakanishi-Lautrup multipliers), and adding to the action the term
\begin{align}
	S_\s{\mathrm{GF}}=\int\!\d t\!\int\! \d^3 \vec{x} \, \left[ b^0 {\cal F}_0 + b^j {\cal F}_j  \right].
	\label{NL.term}
\end{align} 
Clearly, the conditions~\noeq{comov.gauge.infl}  and \noeq{como_trans} are recovered once the $b$-fields are eliminated through their (trivial) equations of motion.

In the temporal sector the only non-vanishing propagator is the mixed one $b^0\delta \phi$. Nevertheless, since there are no interaction vertices involving the $b^0$ field, one can safely set $\delta \phi=0$ everywhere in the effective action. This result will be recovered by functional methods later on in Sect.~\ref{sec.ST} and Appendix~\ref{app.funct}.

On the other hand, the inversion of the metric propagator requires some care. In the ADM formalism it is convenient to exploit the invariance under coordinate reparametrization to express the metric two-point function as 
\begin{align}
\widetilde{S}^{(2)}&=\int\!{\mathrm d}t\!\int\! {\mathrm d}^3 \vec{x} \,\delta h^{ij}\widetilde{\Gamma}^{(2)}_{ij\,mn}\delta h^{mn};&
\widetilde{\Gamma}^{(2)}_{ij\,mn}=\sum_{\alpha=1}^7{\cal O}^{(\alpha)}_{ij\,mn}\theta_\alpha,
\label{quadact}
\end{align}
where the seven bi-tensors ${\cal O}^{(i)}$ are defined as
\begin{align}
	{\cal O}^{(1)}_{ij\,mn}&=\delta_{im}\delta_{jn}+\delta_{jm}\delta_{in};&
	{\cal O}^{(2)}_{ij\,mn}&=\delta_{ij}\delta_{mn};&
	{\cal O}^{(3)}_{ij\,mn}&=\delta_{jn}\frac{\partial_i\partial_m}{\partial^2}+\delta_{in}\frac{\partial_m\partial_j}{\partial^2};\nonumber \\
	{\cal O}^{(4)}_{ij\,mn}&=\delta_{im}\frac{\partial_j\partial_n}{\partial^2}+\delta_{mj}\frac{\partial_i\partial_n}{\partial^2};&
	{\cal O}^{(5)}_{ij\,mn}&=\delta_{mn}\frac{\partial_i\partial_j}{\partial^2};&
	{\cal O}^{(6)}_{ij\,mn}&=\delta_{ij}\frac{\partial_m\partial_n}{\partial^2};\nonumber \\
	{\cal O}^{(7)}_{ij\,mn}&=\frac{\partial_i\partial_m\partial_j\partial_n}{\partial^4}
	\label{bitensors}
\end{align}
and the coefficients $\theta_\alpha$ are differential operators in $\partial_t$ whose explicit form  is given in~\1eq{thethetas}.

Since the $\theta$'s are differential operators, the second variation of the quadratic action with respect to the metric does not coincide with $\widetilde{\Gamma}^{(2)}$ (as it would were the $\theta_\alpha$ algebraic coefficients). Instead, one has (up to total derivative terms)
\begin{align}
	\widetilde{\Gamma}_{ij\,mn}^{'(2)}(\vec{x},t;\vec{z},t')=\frac{\delta^2\widetilde{S}^{(2)\!}}{\delta h^{im}(\vec{x},t)\delta h^{jn}(\vec{z},t')}=\sum_{\alpha=1}^7\delta^3(\vec{z}-\vec{x})\delta(t'-t){\cal O}^{(\alpha)}_{ij\,mn}\Theta_\alpha,
\end{align} 
where the new differential operators $\Theta$ are completely determined by the old ones $\theta$, see~\2eqs{Theta1}{Theta2}.

It is then relatively straightforward to invert the two-point functions in the $b^i$-$\delta g^{jk}$ sector, as it is explicitly shown in Appendix~\ref{app.diff.eq.metric.prop}. In particular, \1eq{grav.prop} shows that the metric propagator is chracterized by two scalar functions  $r_1$ and $r_2$ satisfying a system of coupled differential equations. This system supports also a constant solution, for which $\partial_t r_1=\partial_t r_2=0$. Evidently, the latter has to be identified with the one at the time $t=t_*$ of horizon crossing, where~\cite{Maldacena:2002vr}
\begin{align}
	\dot\rho_*\,{\mathrm e}^{\rho_*}\sim p, 
\end{align}
a $*$ meaning that the corresponding quantity has been evaluated at $t=t_*$. Specifically one finds
\begin{align}
	r_1 = -\frac2{p^2} e^{-\rho_*};\qquad  
	r_2 = \frac2{p^2} e^{-\rho_*}\left(\frac{\dot{\rho}_*^2}{\ddot{\rho}_*}-1\right).
	\label{grav.prop.params}
\end{align}
Notice that $r_2$ is then proportional to the inverse of the slow-roll parameter $\epsilon$ defined in~Eq.\noeq{slowroll} and therefore it can give rise to potentially large effects when gravitons propagate in loops. 

On the other hand, such contributions are expected not to affect physical observables, for the physical propagator for the graviton is controlled by the function $r_1$ alone. To prove this, one observes that all the coefficients $\psi, E, F_i$ and $\gamma_{ij}$ in~\1eq{metric.decomp} can be obtained by applying appropriate projectors to the metric $\delta h_{ij}$; in particular, for the physical graviton $\gamma_{ij}$ we have 
\begin{align}
	\gamma_{ij}&=\frac12P^\gamma_{ij\,mn}\delta h_{mn},
\end{align}
where
\begin{align}
	P^\gamma_{ij\,mn}&=\sum_{\alpha=1}^7c_\alpha{\cal O}^{(\alpha)}_{ij\,mn};& &c_1=-c_2=-c_3=-c_4=c_5=c_6=c_7=1.
\end{align}	
thus giving rise to the physical propagator (with $T$ being the usual time-order product)
\begin{align}
		\langle T[\gamma_{ij}\gamma_{kl}]\rangle&=\frac14P^\gamma_{ij\,mn}P^\gamma_{kl\,pq}\langle \delta h_{mn}\delta h_{pq}\rangle \nonumber \\
		&=\frac14\sum_{\alpha=1}^7r_\alpha P^\gamma_{ij\,mn}P^\gamma_{kl\,pq}{\cal O}^{(\alpha)}_{mn\,pq}=r_1P^\gamma_{ij\,kl}.
\end{align}
We then see that the dependence on $r_2$ has disappeared and one obtains a propagator characterized by the single scalar form factor $r_1$ satisfying the differential equation~\noeq{r1eq}. As a result, since $r_2$ controls the unphysical part of the propagator, one expects that such contributions cancel out in physical amplitudes against corresponding ghost loops contributions.

%

BRST invariance requires finally to add the ghost action, which reads
\begin{align}
S_\s{\mathrm{FPG}} &=  \int\! \d t\int\!\d^3 \vec{x}
\left[ \dot{\varphi} \bar c^0 c^0 - \bar c^i s {\cal F}_i \right].
\label{s.GFpFPG}
\end{align}
The BRST variation of the anti-ghost fields $\bar c^\mu$ is defined by pairing them with the corresponding 
Nakanishi-Lautrup field in a so-called BRST doublet: $s \bar c^\mu = b^\mu$, $s b^\mu = 0$~\cite{Barnich:2000zw,Quadri:2002nh}. This ensures that the sum $S_\s{\mathrm{GF}}+S_\s{\mathrm{FPG}}$ is an exact BRST variation, so that BRST nilpotency ensures that this term decouples from the physical observables of the theory. 

The quadratic part of the ghost action is obtained from~\1eq{s.GFpFPG} when considering the BRST differential in the linearized approximation $s_\s{\!\ell}$, which is obtained by keeping on the right-hand side of~\1eq{brst.full} only terms linear in the fields. Recalling that both $\rho$ and  $\varphi$ are independent of the spatial coordinates, and using the background metric $\widehat{h}$ to raise/lower indices, one obtains 
\begin{align}
\label{brst.lin}
s_\s{\!\ell} \delta {\cal N} &= -\partial_0 c^0;&\quad
s_\s{\!\ell} \delta {\cal N}^i &= \partial^i c^0 - \partial^0 c^i;&\quad s_\s{\!\ell} \delta h_{ij} &= -2 \dot \rho  c^0 \widehat{h}_{ij} - \partial_j c_i - \partial_i c_j;&\quad
s_\s{\!\ell} \delta \phi &= - c^0 \dot{\varphi},
\end{align}
and, accordingly,
\begin{align}
S^{(2)}_\s{\mathrm{FPG}} &=\int\!\d t\!\int\! \d^3 \vec{x}\,\left[ \dot{\varphi} \bar c^0 c^0-\bar c^i \left( \square c_i + \frac{1}{3} \partial_i \partial_j c^j \right)\right].
\end{align}
Notice that there is no mixing term between the temporal and spatial ghosts, so that by inverting this action one immediately gets the  propagators
\begin{align}
	\Delta_{\bar c^0 c^0} &= \dot{\varphi}^{-1};&
	\Delta_{\bar c_i c_j} = \frac{1}{p^2} \left ( \delta_{ij} - \frac{1}{4} \frac{p_i p_j}{p^2} \right).
\end{align}

\section{The Slavnov-Taylor identity}\label{sec.ST}

It is possible to recast the invariance under the BRST symmetry of the classical gauge-fixed action in a functional form through the so-called Slavnov-Taylor (ST) identity.

For that purpose one introduces for each of the fields $\Phi$ with a non-linear BRST variation $s \Phi$ an external source $\Phi^*$, called anti-field. The anti-field $\Phi^*$ has opposite statistics with respect to $\Phi$, whereas its ghost charge (which is a conserved quantum number), gh($\Phi^*$), is related to the ghost charge gh($\Phi$) of the corresponding field through gh($\Phi^*$) = $-1 -$gh($\Phi$). Conventionally setting gh$(c^\mu)=1$, it turns out that the ghost charge of the anti-ghost and all anti-fields, with the exception of $c^*_\mu$, is $-1$; the ghost charge of $c^*_\mu$ is $-2$, whereas all the other fields have ghost number zero. 

Next, anti-fields are coupled to $s\Phi$ in the tree-level vertex functional $\G^{(0)}$ (which has ghost number zero); that is one sets
\begin{align}
\G^{(0)} &= S + S_\s{\mathrm{GF}} + S_\s{\mathrm{FPG}} + S_\s{\mathrm{AF}},
\end{align}
where
\begin{align}
S_\s{\mathrm{AF}} &= \sum_\Phi \int\!\d t\!\int\! \d^3 \vec{x}\, \Phi^* \, s \Phi \, , 
\end{align}
and the sum runs over $\Phi = \{ \delta h_{ij}, \delta {\cal N}, {\cal N}_i, \delta \phi, c_0, c_i \}$. Then 
$\G^{(0)}$ is BRST invariant if and only if one imposes $s \Phi^* = 0$. Notice that we do not introduce a source for $\bar c^\mu$ since its BRST variation is linear.

The invariance of $\G^{(0)}$ under the BRST differential $s$ translates then into the following functional equation, known as the ST identity
\begin{align}
{\cal S}(\G^{(0)}) = \int\!\d t\!\int\! \d^3 \vec{x}\, &\left[
\Gz_{\delta h^{ij*}}\Gz_{\delta h_{ij}}+\Gz_{\delta {\cal{N}}^{*}}\Gz_{\delta {\cal{N}}}+\Gz_{{\cal{N}}^{i*}}\Gz_{{\cal{N}}_i}+\Gz_{\delta\phi^*}\Gz_{\delta\phi}+b_\mu\Gz_{\bar c^\mu}+\Gz_{c^{\mu*}}\Gz_{c_\mu}\right]=0.
\label{SGamma0}
\end{align}
In the equation above we have introduced the notation $\Gz_\Phi=\delta\Gz/\delta\Phi(t,\vec x)$, omitting at the same time the argument of the fields to avoid notational clutter.

Taking functional derivatives of ${\cal S}(\G^{(0)})$ and setting afterwards all fields and anti-fields to zero will generate the complete set of the all-order ST identities of the theory; this is in exact analogy to what happens with the effective action, where taking functional derivatives of $\Gamma$ and setting afterwards all fields to zero generates the 1-PI Green functions of the theory. However, in order to reach meaningful expressions, one needs to keep in mind that: ({\it i}) ${\cal S}(\G^{(0)})$  has ghost charge 1, and ({\it ii}) functions with non-zero ghost charge vanish, since the latter is a conserved quantity. Thus, in order to extract non-zero identities from~\1eq{SGamma0}, one needs to differentiate the latter with respect to a combination of fields, containing either one ghost field, or two ghost fields and one anti-field. The only exception to this rule is when differentiating with respect to a ghost anti-field, which needs to be compensated by three ghost fields. 

The dependence of  $\G$ on the gauge-fixing and the anti-ghost fields is controlled
by the $b$-equations and the anti-ghost equations. They are (we omit the space-time arguments for simplicity)
\begin{align}
 \Gz_{b^0} &= \delta \phi,&  &{\mbox{(temporal {\it b}-equation)}} \nonumber \\
 \Gz_{b^i} &= \partial^k \delta h_{ki} - \frac{1}{3} \partial_i \delta h^k_k ,&
&{\mbox{(spatial {\it b}-equation)}} \nonumber \\
 \Gz_{\bar c^0} &= - \Gz_{\delta \phi^*} , &
&{\mbox{(temporal anti-ghost equation)} } \nonumber \\
 \Gz_{\bar c^i} &=  \partial^k \Gz_{\delta h^{ki*}} -
\frac{1}{3} \partial_i \Gz_{\delta h^{k*}_k}&
&{\mbox{(spatial anti-ghost equation)} } 
\end{align}
The temporal anti-ghost equation states that the dependence of the vertex functional on the temporal anti-ghost field
happens only via the combination $\delta \phi^{*'} =  \delta \phi^* - \bar c^0$. In a similar way the
spatial anti-ghost equation guarantees that the dependence of the vertex functional on the spatial
anti-ghosts $\bar c^i$ is through the combination 
$\delta h^{ij*'} = \delta h^{ij*} + \partial^i \bar c^j - \frac{1}{3} \partial_\ell \bar c^\ell \delta h^{ij} $ only.

In classical cosmology computations it is customary to eliminate the lapse function and shift vector through their equations of motion. In our formulation this can be achieved by considering the generating functional $\Gt$ for diagrams which are one-particle reducible with respect to lapse and shift propagators and one-particle irreducible with respect to all other fields; details of the algebraic construction of $\Gt$ are reported in Appendix~\ref{app.funct}.

It turns out that the effective action $\Gt$ satisfies the following ST identity:
\begin{align}
\label{eff.sti}
{\cal S}(\widetilde \G) = \int\!\d t\!\int\! \d^3 \vec{x}\, \left[
\Gt_{\delta h^{ij*}}\Gt_{\delta h_{ij}} 
+ \Gt_{\delta\phi^*} \Gt_{\delta\phi}+ b^\mu \Gt_{\bar c^\mu} + 
\Gt_{c^{\mu *} } \Gt_{c_\mu } 
\right]  = 0.
\end{align}
 This is the basic functional relation, valid in any gauge, from which we will start deriving the consistency conditions of the inflationary theory. It is worthwhile to notice that, being based on the diffeomorphism invariance only, this is the most natural formulation that is supposed to hold true even when loop corrections are  included (provided that one can consistently make sense of the UV and IR divergences arising in graviton loops, something which is far beyond the scope of the present paper).

The relevant consistency conditions embodied in the ST identity~\noeq{eff.sti} are of two types. 

If one takes a derivative with respect to the ghost field $c$ and then with respect to a combination of zero ghost charge fields other than the multiplier $b$ and afterwards set all fields and external sources to zero, one gets relations that are valid in any gauge (the gauge dependence being controlled by the $b$-term in the ST identity). 

On the other hand, if we take a derivative of~\1eq{eff.sti} with respect to the combination $cb$ and then set $c$ to zero, one obtains a functional identity encoding the gauge dependence of the vertex functional. This is further discussed in Appendix~\ref{app.gauge.indip.mast}.


\section{Spacetime symmetries in BRST language}\label{sec.maldacena}

In order to proceed further, we need to elaborate on the relation between the BRST symmetry of the action and its (classical) invariance under diffeomorphism, dilatation and special conformal transformations.  To this end, we will adopt an explicit parametrization of the metric $h_{ij}$; specifically we write~\cite{Maldacena:2002vr}
\begin{align}
	h_{ij}=\widehat{h}_{ij}+e^{2\rho}\delta h_{ij},
\end{align}
where 
\begin{align}
	\widehat{h}_{ij}&=e^{2\rho}e^{2\zeta}\delta_{ij};&
	\delta h_{ij}&=e^{2\zeta}\left(\gamma_{ij}+\frac12\gamma_i^k\gamma_{kj}+\cdots\right);&
	\partial^i\gamma_{ij}=\gamma_i^i=0.
	\label{malconv}
\end{align}
Within this parametrization, the ST identity~\noeq{eff.sti} reads then
\begin{align}
	{\cal S}(\widetilde \G^{(0)}) = \int\!\d t\!\int\! \d^3 \vec{x}\, \left[\Gzt_{\zeta^*} \Gzt_{\zeta}+
\Gzt_{\gamma^{ij*}} \Gzt_{\gamma_{ij}} 
+ \cdots\right]=0,
\label{eff.sti.mald}
\end{align} 
where $\widetilde{\Gamma}^{(0)}$ is the tree-level version of the generating functional introduced in the previous section, and the dots indicate terms containing the $b$ or the ghost field.

As already discussed, if we are interested only in gauge independent ST identities, we can take a derivative with respect to the ghost field and drop all gauge variant terms, to obtain the relation
\begin{align}
	\int\!\d t\!\int\! \d^3 \vec{x}\,\left[\widetilde{\Gamma}^{(0)}_{c^\ell\zeta^*}\widetilde{\Gamma}^{(0)}_\zeta+\widetilde{\Gamma}^{(0)}_{c^\ell\gamma^{*ij}}\widetilde{\Gamma}^{(0)}_{\gamma_{ij}}\right]=0.
	\label{M-master}
\end{align} 
We will refer to~\1eq{M-master} as the `master' consistency relation; in fact we will show that the known consistency conditions related to diffeomorphism, dilation or special conformal transformations invariance, are all descendant of the basic identity~\noeq{M-master}. However, before we can do that, we need first to explicitly derive the BRST transformations of the $\zeta$ and $\gamma$ fields. 

\subsection{BRST transformations of $\zeta$ and $\gamma$}

In general it is not possible to derive closed expressions for the BRST variations $s\zeta$ and $s\gamma$; however a recursive procedure can be devised that allows to grade them according to the number of graviton fields $\gamma$, the resulting expressions being however valid to all orders in the $\zeta$ field. 

We start by taking a BRST variation of the metric $h_{ij}$ in the representation~\noeq{malconv}, yielding  up to second order in $\gamma$
\begin{align}
	sh_{ij}&=2(s\zeta)e^{2\rho}e^{2\zeta}\left(\delta_{ij}+\gamma_{ij}+\frac12\gamma_i^k\gamma_{kj}\right)+\frac12e^{2\rho}e^{2\zeta}\left[2s\gamma_{ij}+\left(s\gamma_i^k\right)\gamma_{kj}+\gamma_i^ks\left(\gamma_{kj}\right)\right].
	\label{sh-1}
\end{align}
On the other hand,~\1eq{brst.full} gives for the same BRST variation the result
\begin{align}
	sh_{ij}&=-c^m\partial_m\left[e^{2\rho}e^{2\zeta}\left(\delta_{ij}+\gamma_{ij}+\frac12\gamma_i^k\gamma_{kj}\right)\right]-e^{2\rho}e^{2\zeta}\left(\delta_{im}+\gamma_{im}+\frac12\gamma_i^k\gamma_{km}\right)\partial_j c^m\nonumber \\
	&-e^{2\rho}e^{2\zeta}\left(\delta_{jm}+\gamma_{jm}+\frac12\gamma_j^k\gamma_{km}\right)\partial_i c^m.
	\label{sh-2}
\end{align} 

Tracing both equations, and equating the zeroth order terms in the graviton field gives
\begin{align}
	(s\zeta)_{0}=-c^m\partial_m\zeta-\frac13\partial_m c^m.
	\label{sz0}
\end{align}
Again by equating terms of order zero (but this time without taking the trace), and using the result above, we get
\begin{align}
	(s\gamma_{ij})_{0}=\frac23\delta_{ij}\partial^m c_m-\partial_jc_i-\partial_ic_j.
	\label{sg0}
\end{align}
The grade 1 BRST transformations can be obtained in exactly the same way. In particular, from tracing the relations~\noeq{sh-1} and~\noeq{sh-2} and keeping the terms with one graviton fields, we obtain
\begin{align}
	(s\zeta)_{1}=-\frac13\gamma^{im}\partial_ic_m+\frac16\gamma^{im}(\partial_i c_m+\partial_m c_i)=0,
	\label{sz1}
\end{align} 
while, finally, using the above results together with \2eqs{sh-1}{sh-2} will give
\begin{align}
	(s\gamma_{ij})_{1}=-c^m\partial_m\gamma_{ij}+\frac12\gamma_{im}\left(\partial^mc_j-\partial_jc^m\right)+\frac12\gamma_{jm}\left(\partial^mc_i-\partial_ic^m\right).
	\label{sg1}
\end{align} 

This is as far as we can go, without considering in the expansion~\noeq{malconv} pieces containing three $\gamma$ fields; it will however turn out to be enough for the ensuing analysis.

\subsection{Dilatation and special conformal transformations}

At the classical level invariance under BRST stems from gauge invariance of the original action; in a gravitational theory therefore it arises from diffeomorphism (or coordinate transformation) invariance. Of particular relevance for the discussion that will follow, are the invariance under dilatations~\cite{Maldacena:2002vr} and special conformal transformations\footnote{These transformations of the spatial coordinates are related to certain isometries of de Sitter space (besides translations, rotations and dilatations) taken at large times~\cite{Creminelli:2012ed}.}~\cite{Creminelli:2012ed}, which are obtained by carrying out the following replacements of the ghost field:
\begin{align}
	c_m&\to\lambda x_m;& &\mathrm{dilatations},\label{dil}\\
	c_m&\to b^i{\cal M}_{im}(\vec{x}\,)\equiv b^i[-\delta_{im}\vec{x}^{\,2}+2x_ix_m];& &\mathrm{special\ conformal\ transformations.}\label{sct}
\end{align}
It is then immediate to realize from~\1eq{sz0} that in both cases one has
\begin{align}
	(s\gamma)_{0}=0.
	\label{zerograde}
\end{align}
As a consequence:
\begin{itemize}
	\item The two-point function $\widetilde\Gamma^{(0)}_{c^\ell\gamma^*}$ which is controlled by the antifield combination $\gamma^*(s\gamma)_{0}$ vanishes, along with all other $n$-point functions $\widetilde\Gamma^{(0)}_{c^\ell\gamma^*\zeta_1\cdots\zeta_{n-2}}$ (with $n>2$)\footnote{These functions vanish in general for the classical case. This is because~\1eq{sg0} contains no coupling to the $\zeta$ field and the BRST variations found are all order expressions in the $\zeta$ field.}. 
	\item The three-point function $\widetilde\Gamma^{(0)}_{c^\ell\zeta^*\gamma}$ that is controlled by the antifield combination $\zeta^*(s\zeta)_{1}$ vanishes.
 \end{itemize}  
 
As we will see, dilatations and special conformal transformations determine the leading and next to leading order in the squeezed limit expansion of a 1-PI function (or correlator) involving a soft scalar; then~\1eq{zerograde} ensures that there are no graviton contributions for purely scalar functions (or correlators) at the leading and next to leading order.    

\section{\label{sec:cosmological.sti}Cosmological ST identities}

We are now in a position to work out the ST identities dictated by dilatation, diffeomorphism and special conformal invariance, ultimately showing that they coincide with the consistency relations worked out in~\cite{Maldacena:2002vr} and~\cite{Creminelli:2012ed}.
Before doing this it is necessary however to discuss the standard formalism for computing correlation functions in cosmological models, and how these functions are related to the 1-PI functions naturally appearing in our approach.

\subsection{In-in formalism}

Cosmological correlation functions are computed in the
so-called ``in-in'' formalism~\cite{Weinberg:2005vy}.
This formalism allows to evaluate expectation values of products
of fields at a fixed time by imposing conditions on the fields
at very early times  (in the present paper chosen to be
those corresponding to Bunch-Davies states~\cite{Maldacena:2002vr}).

The correlators of products of fields at the given time $t$ are collected
into a functional $W(t)$, defined as the generating functional
of the ``in-in'' correlators.
Namely by collectively denoting the quantum fields by $\delta \phi(t,\vec{x})$
and by $J(\vec{x})$ the corresponding source,
$W(t)$ is given by
\begin{equation}
	W(t) = \sum_{N=2}^{\infty} \int\! {\mathrm d}^3x_1 \cdots\! \int\! {\mathrm d}^3x_N \,\frac{1}{N!} \langle \delta \phi(t,\vec{x}_1) \dots\delta \phi(t,\vec{x}_N) \rangle J(\vec{x}_1) \dots J(\vec{x}_N).
	\label{eq.in.1}
\end{equation}
The path-integral generating the correlators $\langle \delta \phi(t,\vec{x}_1) \dots\delta \phi(t,\vec{x}_N) \rangle $ is written in the Schwinger-Keldysh formalism~\cite{Weinberg:2005vy}. One needs to double the fields of the theory (we call them left- and right fields, in agreement with the notation of~\cite{Weinberg:2005vy}). Then the path-integral generating $W(t)$ can be written as ($\delta \pi$ denote the conjugate momenta to $\delta \phi$)
\begin{eqnarray}
	e^{W(t)/g} & = & \int  \prod \delta \phi_\s{\mathrm L} \prod \delta \pi_\s{\mathrm L}\prod \delta \phi_\s{\mathrm R} \prod \delta \pi_\s{\mathrm R}\nonumber \\ 
	& & \exp \left ( -i \int_{-\infty}^t dt' \, \frac{1}{g} 
	\widetilde L[\delta \phi_\s{\mathrm L}, \delta \pi_\s{\mathrm L}; t']\right ) \times \exp \left ( +i \int_{-\infty}^t dt' \, \frac{1}{g} 
	\widetilde L[\delta \phi_\s{\mathrm R}, \delta \pi_\s{\mathrm R}; t'] \right ) \nonumber \\
	& & \times \prod \delta[\delta \phi_\s{\mathrm L}(t) - \delta \phi_\s{\mathrm R}(t)]
	\times \prod \delta[\delta \pi_\s{\mathrm L}(t) - \delta \pi_\s{\mathrm R}(t)]
	\times \exp\left({\frac{1}{g} \int\mathrm{d}^3 \vec{x} \,  \delta \phi(\vec{x},t) J(\vec{x})}\right) \nonumber \\
	& & \times \Psi_{\mathrm{vac}}[\delta \phi_\s{\mathrm L}(-\infty)] \Psi_{\mathrm{vac}}[\delta \phi_\s{\mathrm R}(-\infty)] 
\label{eq.path.int}
\end{eqnarray}
where $\widetilde L$ is the Lagrangian and $\Psi_\mathrm{vac}[\delta \phi]$ is the wave function of the vacuum. In the above equation we have re-installed a coupling constant $g$.

The classical approximation is recovered in the limit $g \rightarrow 0$, since by standard power-counting arguments one can show that the $g$-dependence of $W(t)$ is at loop $n$ is $g^{-n}$. The tree graphs are then recovered in the limit $g \rightarrow 0$. In this limit the path-integral (\ref{eq.path.int}) is dominated by the stationary points of the Lagrangian, namely $\delta \phi_\s{\mathrm L} = \delta \phi_\s{\mathrm R} = \delta \phi_{cl}$, and $\delta \pi_\s{\mathrm L} = \delta \pi_\s{\mathrm R} = \delta \pi_{cl}$, and the classical solutions satisfy the prescribed Bunch-Davies conditions at $t= -\infty$. At the stationary point the action integrals cancel and one recovers the well-known result~\cite{Maldacena:2002vr} that at the classical level the ``in-in'' correlation functions are computed by taking the product of the fields obtained as the solutions of the classical field equations with the given free-field initial conditions.

\subsection{ST identity for $W(t)$ and its Legendre transform}

In Eq.(\ref{eq.path.int}) 
$\widetilde L$ is the gauge-fixed Lagrangian. Since the integration in the action integrals is limited to $t$, only invariance under spatial BRST transformations (i.e. those obtained by setting $c^0=0$) holds for $\widetilde L$. Assuming that the vacuum wave function $\Psi_\mathrm{vac}$ 
	and the integral measure are also invariant under spatial diffeomorphisms,
	then $W(t)$, as defined in~\1eq{eq.path.int}, satisfies the ST identity ($J_
	\Phi$ denotes the source of the field $\Phi$)
\begin{align}
	\int \!\d^3x &\left[  \frac{\delta W(t)}{\delta (\delta h^*_{{ij}}(t,\vec{x}))} J_{\delta h_{ij}}(\vec{x})   +\frac{\delta W}{\delta (\delta \phi^*(t,\vec{x}))} J_{\delta \phi}(\vec{x})\right. \nonumber \\
	& +\left. 
	   \frac{\delta W}{\delta J_{b^\mu}(t,\vec{x})} J_{\bar c^\mu}(\vec{x}) +
	   \frac{\delta W}{\delta c^{\mu*}(t,\vec{x})} J_{c^\mu}(\vec{x}) \right] = \Delta(c^0), 
	   \label{Wt.st} 
\end{align}
where $\Delta(c^0)$ is a breaking term vanishing at $c^0 = \frac{\delta W}{\delta J_{c^0}} = 0$.

The ST identity for $W(t)$ in Eq.(\ref{Wt.st}) embodies relations between
connected Green's functions involving at least one insertion of the BRST transformation of the fields. Therefore it cannot be directly used to obtain
 the consistency conditions between the correlators of the scalar and graviton fields in the squeezed limit.
 
The appropriate way to recover such consistency conditions is  
to carry out a Legendre transformation according to
\begin{align}
	\widetilde \G^{(0)}(t) &= W(t)  + \int\!\mathrm{d}^3x\, J(\vec{x}) \, \Phi(t,\vec{x});&
	\Phi(t,\vec{x}) &= \frac{\delta W(t)}{\delta J(\vec{x})},
	\label{legendre.transf}
\end{align}
where the superscript ``(0)'' reminds us that we are working at the tree (classical) level. The two-point 1-PI amplitude is then minus the inverse of the bi-spectrum, e.g., for the scalar correlator 
\begin{equation}
	\Gzt(t)_{\zeta(t,\vec{x}_1) \zeta(t,\vec{x}_2)} = - \langle \zeta(t,\vec{x}_1) \zeta(t,\vec{x}_2) \rangle^{-1},
	\label{2point}
\end{equation}
(a similar relation holds for the graviton correlator).

Then if one sets $c^0=0$, which is possible at the classical level, where one does not need an additional source to define the composite operator $\Delta(c^0)$, one obtains the basic ST identity for the fixed time 1-PI generating functional $\widetilde \G^{(0)}(t)$, which we write in the parameterization of the metric given in Eq.(\ref{malconv}) :
\begin{align}
	{\cal S}(\widetilde \G^{(0)}(t)) = \int\! \d^3 \vec{x}\, \Big [ & \Gzt(t)_{\zeta^*} \Gzt(t)_{\zeta}+ 
\Gzt(t)_{\gamma^{ij*}} \Gzt(t)_{\gamma_{ij}} \nonumber \\
& + b^\mu \Gzt(t)_{\bar c^\mu} + \Gzt(t)_{c^*_\mu} \Gzt(t)_{c^\mu} \Big ]_{c^0=0}=0 \, .
\label{eff.sti.mald.at.time.t}
\end{align} 
Notice that~\1eq{brst.ghost} implies that $s c^0$ vanishes at $c^0=0$.

By exploiting the decomposition of the connected amplitudes in terms of the 1-PI vertices, one can easily obtain the desired consistency conditions between in-in correlators. 

The only property used in the ensuing derivation is the fulfilment of the ST identity~\1eq{eff.sti.mald.at.time.t}. We remark that one does not need to explicitly know the 1-PI vertices generated by $\Gzt(t)$ (apart from some regularity assumptions in the squeezed limit where one of the momenta is much smaller than the others, a condition that will be checked against explicit computations of the connected correlators available in the literature). In particular it is important to realize that $\Gzt(t)$ does not coincide with the classical action, as is already apparent from the two-point function~\noeq{2point}.

In what follows we will omit the time-dependence on $\Gzt$ in order to avoid notational clutter.

\subsection{Dilatations ST identities}

\subsubsection{Three scalars (one soft)}

The relation between the connected and 1-PI Green function for the case of three scalars is
\begin{align}
	\langle\zeta(\vec{x}\,')\zeta(\vec{y}\,')\zeta(\vec{z}\,')\rangle&=
	\int\! \d^3 \vec{x}\!\int\! \d^3 \vec{y}\!\int\! \d^3 \vec{z}\, \langle\zeta(\vec{x}\,')\zeta(\vec{x}\,)\rangle\langle\zeta(\vec{y}\,')\zeta(\vec{y}\,)\rangle\langle\zeta(\vec{z}\,')\zeta(\vec{z}\,)\rangle\Gzt_{\zeta\zeta\zeta}(\vec{x},\vec{y},\vec{z}\,).
	\label{con1PI-3zeta}
\end{align}
The ST identity that controls the 1-PI function $\Gzt_{\zeta\zeta\zeta}$ can be obtained by taking two derivatives with respect to the scalar field $\zeta$ of the master equation~\noeq{M-master}, and setting afterwards all fields to zero. Then, using the fact that the tadpole contributions $\widetilde\Gamma^{(0)}_\zeta$ and $\widetilde\Gamma^{(0)}_{\gamma_{ij}}$ vanish, we obtain in the dilatation case
\begin{align}
	\int\!\d^3 \vec{w}\,&\left[\Gzt_{\zeta c^\ell\zeta^*}(\vec{y},\vec{x},\vec{w}\,)\Gzt_{\zeta\zeta}(\vec{w},\vec{z}\,)+
		\Gzt_{\zeta c^\ell\zeta^*}(\vec{z},\vec{x},\vec{w}\,)\Gzt_{\zeta\zeta}(\vec{w},\vec{y}\,)+
		\Gzt_{c^\ell\zeta^*}(\vec{x},\vec{w}\,)\Gzt_{\zeta\zeta\zeta}(\vec{w},\vec{y},\vec{z}\,)\right.\nonumber \\
		&\left.+\Gzt_{c^\ell\gamma^*_{ij}}(\vec{x},\vec{w}\,)\Gzt_{\gamma^{ij}\zeta\zeta}(\vec{w},\vec{y},\vec{z}\,)\right]
	=0.
	\label{3sf}
\end{align}

For the dilatation case the last term vanishes at leading order, and therefore only the 2- and 3-point functions $\Gzt_{c^\ell\zeta^*}$ and $\Gzt_{\zeta c^\ell\zeta^*}$ need to be considered. The latter are both generated by the term $\int\zeta^*(s\zeta)_{0}$ appearing in the antifield action; in particular, at the classical level, one has the results
\begin{align}
	\Gzt_{c^\ell\zeta^*}(\vec{x},\vec{w}\,)&=-\frac13\partial_{w^\ell}\,\delta(\vec{x}-\vec{w}\,),\label{Gczeta*}\\
	\Gzt_{\zeta c^\ell\zeta^*}(\vec{y},\vec{x},\vec{w}\,)&=-\delta(\vec{y}-\vec{w}\,)\partial_{w^\ell}\delta(\vec{x}-\vec{w}\,),
\end{align}
which inserted back into~\1eq{3sf} gives 
\begin{align}
	\frac13\partial_{x^\ell}\Gzt_{\zeta\zeta\zeta}(\vec{x},\vec{y},\vec{z}\,)-\partial_{x^\ell}\delta(\vec{x}-\vec{y}\,)\Gzt_{\zeta\zeta}(\vec{x},\vec{z}\,)-\partial_{x^\ell}\delta(\vec{x}-\vec{z}\,)\Gzt_{\zeta\zeta}(\vec{x},\vec{y}\,)=0.
	\label{confgeneral}
\end{align}

As already said, for a dilatation transformation one has $c^\ell\sim x^\ell$, so that to get back the effect of such transformation on our Green functions we need to multiply by $x^\ell$ and integrate over $\mathrm{d}^3\vec{x}$; we then obtain
\begin{align}
	\int\!\mathrm{d}^3\vec{x}\,\Gzt_{\zeta\zeta\zeta}(\vec{x},\vec{y},\vec{z}\,)=\left(6+\vec{y}\spr\vec{\partial}_y+\vec{z}\spr\vec\partial_z\right)\Gzt_{\zeta\zeta}(\vec{y},\vec{z}\,).
	\label{3scalxspace}
\end{align}
Notice the automatic appearance of the squeezed limit as the $\d^3\vec{x}$ integral ensures that the corresponding scalar field is inserted with zero momentum.

To see this explicitly, let's now Fourier transform everything to momentum space. Using translational invariance one can set to zero say the $\vec{z}$ coordinate; using then momentum conservation $\vec{k}_2=-\vec{q}-\vec{k}_1$, so that, setting $\vec{k}_1=\vec{k}$, the first term in the equation above reads\footnote{Here and in the entire ensuing analysis we will omit from all correlation functions the overall momentum conserving delta function $(2\pi)^3\delta(\sum\vec{k})$.}
\begin{align}
	\int\!\mathrm{d}^3\vec{x}\,\Gzt_{\zeta\zeta\zeta}(\vec{x},\vec{y},\vec{z}\,)&=\int\!\mathrm{d}^3\vec{x}\,\!\int\!\frac{\d^3\vec{q}}{(2\pi)^3}\,\,e^{-i\vec{q}\cdot \vec{x}}\!\int\!\frac{\d^3\vec{k}}{(2\pi)^3}\,e^{-i\vec{k}\cdot \vec{y}}\,\Gzt_{\zeta\zeta\zeta}(\vec{q},\vec{k},-\vec{q}-\vec{k}\,)\nonumber \\
	&=\int\!\frac{\d^3\vec{k}}{(2\pi)^3}\,e^{-i\vec{k}\cdot \vec{y}}\,\Gzt_{\zeta\zeta\zeta}(0,\vec{k},-\vec{k}\,).
\end{align}

For the second term, again making use of translational invariance and the resulting momentum conservation, one finds
\begin{align}
	\left(\vec{y}\spr\vec{\partial}_y+\vec{z}\spr\vec\partial_z\right)\Gzt_{\zeta\zeta}(\vec{y},\vec{z}\,)&=\vec{y}\spr\vec{\partial}_y\int\!\frac{\d^3\vec{k}}{(2\pi)^3}\,e^{-i\vec{k}\cdot \vec{y}}\,\Gzt_{\zeta\zeta}(\vec{k},-\vec{k}\,)\nonumber \\
	&=-\int\!\frac{\d^3\vec{k}}{(2\pi)^3}\,e^{-i\vec{k}\cdot \vec{y}}\left(3+\vec{k}\spr\vec{\partial}_k\right)\Gzt_{\zeta\zeta}(\vec{k},-\vec{k}\,).
\end{align}
Therefore, in momentum space \1eq{3scalxspace} reads
\begin{align}
	\Gzt_{\zeta\zeta\zeta}(0,\vec{k},-\vec{k}\,)=\left(3-\vec{k}\spr\vec{\partial}_k\right)\Gzt_{\zeta\zeta}(\vec{k},-\vec{k}\,).
	\label{res1}
\end{align}
This relation can be further simplified at horizon crossing, where $\Gzt_{\zeta\zeta}(\vec{k},-\vec{k}\,)\sim k^{3-n_s}$~\cite{Maldacena:2002vr}; thus we obtain the final result
\begin{align}
	\Gzt_{\zeta\zeta\zeta}(0,\vec{k},-\vec{k}\,)=n_s\Gzt_{\zeta\zeta}(\vec{k},-\vec{k}\,).
	\label{malda1-1PI}
\end{align}

Before moving on, notice the following. The absence of singularities in the 1-PI three-point function $\Gzt_{\zeta\zeta\zeta}(\vec{q},\vec{k}_1,\vec{k_2}\,)$ in the squeezed limit $\vec{q}\to0$ can be checked by direct inspection of the expressions given in \cite{Maldacena:2002vr}, once one recovers the corresponding 1-PI amplitude from the connected one by multiplying the latter by the appropriate (inverse) external propagators. Thus one can write in this limit\footnote{Notice however that, in general, there might be singularities in the effective action $\Gzt$. This is because it contains diagrams that are one-particle reducible with respect to the lapse and the shift vector; thus, once one eliminates these auxiliary fields through their equations of motion, the appearance of $1/\partial^2$ terms in $\psi$ in \1eq{lapse.shift.vec.eom} via the dependence on $\chi$ might in principle yield singularities at zero momentum in some kinematical soft configurations.}
\begin{align}
	\Gzt_{\zeta\zeta\zeta}({\vec q},\vec{k}_1,\vec{k_2}\,)=\Gzt_{\zeta\zeta\zeta}(0,\vec{k},-\vec{k}\,)+\left.\vec{q}\spr\vec{\partial}\,\Gzt_{\zeta\zeta\zeta}({\vec q},\vec{k}_1,\vec{k_2}\,)\right\vert_{\vec{q}=0}+{\cal O}(q^2).
	\label{texp}
\end{align}
Then,~\1eq{3scalxspace} through~\noeq{malda1-1PI} show that the leading order term in this expansion is completely determined by the dilatation transformations $c^m\to\lambda x^m$; in addition, as already anticipated, the vanishing of the BRST variation of the (zero grade) graviton field under this transformation, \1eq{zerograde}, implies that there will be no graviton contribution at this order. We will see in Sect.~\ref{SCTs} that the next to leading order in the expansion~\noeq{texp} is determined by the special conformal transformations of~\cite{Creminelli:2012ed}.   

Coming back to~\1eq{malda1-1PI}, as a final step let us recast this result in terms of connected Green functions in the squeezed limit $\vec{q}\to0$, evaluating the leading term of the resulting amplitude. In momentum space~\1eq{con1PI-3zeta} reads
\begin{align}
	\langle\zeta(\vec{q}\,)\zeta(\vec{k}_1)\zeta(\vec{k}_2)\rangle&=
	\langle\zeta(\vec{q}\,)\zeta(-\vec{q}\,)\rangle\langle\zeta(\vec{k}_1)\zeta(-\vec{k}_1)\rangle\langle\zeta(\vec{k}_2)\zeta(-\vec{k}_2)\rangle\Gzt_{\zeta\zeta\zeta}(\vec{q},\vec{k}_1,\vec{k}_2).
	\label{con1PI-3zeta.momspace}
\end{align}
Now, the 2-point correlator $\langle\zeta(\vec{q}\,)\zeta(-\vec{q}\,)\rangle$ diverges as $q^{-3}$ in this limit. On the other hand, we know that the 1-PI Green function $\Gzt_{\zeta\zeta\zeta}(\vec{q},\vec{k}_1,-\vec{k}_1-\vec{q}\,)$ is smooth in the same limit, and therefore we can safely replace $\Gzt_{\zeta\zeta\zeta}(\vec{q},\vec{k}_1,-\vec{k}_1-\vec{q}\,)$  with its value at $\vec{q}=0$. Thus, using  the result~\1eq{malda1-1PI} we finally obtain\footnote{Within our convention on the Legendre transform one has that the inverse of the propagator is {\it minus} the two-point 1-PI function: ${\cal W}(t)_{J_\Phi J_\Phi}\widetilde\Gamma^{(0)}_{\Phi\Phi}=-1$, with ${W}(t)_{J_\Phi J_\Phi}\equiv\langle\Phi\Phi\rangle$.}
\begin{align}
	\langle\zeta(\vec{q}\,)\zeta(\vec{k}_1)\zeta(\vec{k}_2)\rangle \underset{\vec{q}\to 0}{\sim} -n_s\langle\zeta(\vec{q}\,)\zeta(-\vec{q}\,)\rangle\langle\zeta(\vec{k}_1)\zeta(-\vec{k}_1)\rangle.
	\label{malda1-con}
\end{align} 

\subsubsection{One soft scalar and two gravitons}

The derivation of this ST identity is very close to what we have described in the previous case of three scalar fields. One starts from the 3-point correlator
\begin{align}
	\langle\zeta(\vec{x}\,')\gamma^{\lambda_1}(\vec{y}\,')\gamma^{\lambda_2}(\vec{z}\,')\rangle&=\int\! \d^3 \vec{x}\!\int\! \d^3 \vec{y}\!\int\! \d^3 \vec{z}\, \langle\zeta(\vec{x}\,')\zeta(\vec{x}\,)\rangle\sum_{\lambda,\lambda'=\pm}\epsilon^{\lambda'}_{ab}(\vec{y}\,')\epsilon^\lambda_{mn}(\vec{y}\,)\langle\gamma^{\lambda'}(\vec{y}\,')\gamma^\lambda(\vec{y}\,)\rangle\nonumber \\
	&\times\sum_{\lambda,\lambda'=\pm}\epsilon^{\lambda'}_{cd}(\vec{z}\,')\epsilon^\lambda_{rs}(\vec{z}\,)\langle\gamma^{\lambda'}(\vec{z}\,')\gamma^\lambda(\vec{z}\,)\rangle\Gzt_{\zeta\gamma^{mn}\gamma^{rs}}(\vec{x},\vec{y},\vec{z}\,)\epsilon_{ab}^{\lambda_1}(\vec{y}\,')\epsilon_{cd}^{\lambda_2}(\vec{z}\,'),
	\label{con1PI-1zeta2gamma}
\end{align}
with the polarization tensors normalized according to $\epsilon^{\lambda\,ij}\epsilon_{ij}^{\lambda'}=2\delta^{\lambda\lambda'}$ and such that \mbox{$\epsilon^\lambda_{ii}=\partial^i\epsilon^\lambda_{ij}=0$}. 

The ST identity governing the behavior of the three-point 1-PI function $\Gzt_{\zeta\gamma\gamma}$ is then obtained by taking two derivatives with respect to a graviton field of the master equation~\noeq{M-master}, and setting all fields to zero afterwards. This gives
\begin{align}
	\int\!\d^3 \vec{w}\,&\left[\Gzt_{\gamma^{mn}c^\ell\gamma_{ij}^*}(\vec{y},\vec{x},\vec{w})\Gzt_{\gamma^{ij}\gamma^{rs}}(\vec{w},\vec{z})+
	\Gzt_{\gamma^{rs}c^\ell\gamma_{ij}^*}(\vec{z},\vec{x},\vec{w})\Gzt_{\gamma^{ij}\gamma^{mn}}(\vec{w},\vec{y})\right.\nonumber \\
	&\left.+\Gzt_{c^\ell\zeta^*}(\vec{x},\vec{w}\,)\Gzt_{\zeta\gamma^{mn}\gamma^{rs}}(\vec{w},\vec{y},\vec{z}\,)+\Gzt_{c^\ell\gamma_{ij}^*}(\vec{x},\vec{w})\Gzt_{\gamma^{ij}\gamma^{mn}\gamma^{rs}}(\vec{w},\vec{y},\vec{z}\,)\right]=0,
	\label{1s2g}
\end{align}
with the last term in the sum being zero (at tree-level) under dilatation transformations.
At tree-level we have
\begin{align}
	\Gzt_{\gamma_{ab}c^\ell\gamma_{ij}^*}(\vec{y},\vec{x},\vec{w}\,)&=-\delta(\vec{x}-\vec{w})\delta_{ai}\delta_{bj}\partial_{w^\ell}\delta(\vec{w}-\vec{y})\nonumber \\
	&+\frac12\delta_{ai}\delta_{bm}\delta(\vec{w}-\vec{y})\left[\delta_j^\ell\partial_{w^m}\delta(\vec{x}-\vec{w})-\delta^{m\ell}\partial_{w^j}\delta(\vec{x}-\vec{w})\right]\nonumber \\
	&+\frac12\delta_{aj}\delta_{bm}\delta(\vec{w}-\vec{y})\left[\delta_i^\ell\partial_{w^m}\delta(\vec{x}-\vec{w})-\delta^{m\ell}\partial_{w^i}\delta(\vec{x}-\vec{w})\right].
	\label{G_gcg*}
\end{align}
We now insert the equation above together with~\1eq{Gczeta*} back into~\1eq{1s2g}; next we multiply by $x^\ell$ and integrate in ${\mathrm d}^3\vec{x}$ to get, in a completely analogous way to the three scalar case\footnote{Notice that under the dilatation transformation the only non-zero term in~\1eq{G_gcg*} is the first one.} 
\begin{align}
	\int\!\mathrm{d}^3\vec{x}\,\Gzt_{\zeta\gamma^{mn}\gamma^{rs}}(\vec{x},\vec{y},\vec{z}\,)=\left(6+\vec{y}\spr\vec{\partial}_y+\vec{z}\spr\vec{\partial}_z\right)\Gzt_{\gamma^{mn}\gamma^{rs}}(\vec{y},\vec{z}\,),
\end{align}
and therefore, in momentum space, 
\begin{align}
	\Gzt_{\zeta\gamma^{mn}\gamma^{rs}}(0,\vec{k},-\vec{k})&=\left(3-\vec{k}\spr\vec{\partial}_k\right)\Gzt_{\gamma^{mn}\gamma^{rs}}(\vec{k},-\vec{k}).
	\label{res2}
\end{align}

At horizon crossing, one has $\Gzt_{\gamma\gamma}(\vec{k},-\vec{k})\sim k^{3-n_t}$~\cite{Maldacena:2002vr}; thus, we get the final relation 
\begin{align}
	\Gzt_{\zeta\gamma^{mn}\gamma^{rs}}(0,\vec{k},-\vec{k})&=n_t\Gzt_{\gamma^{mn}\gamma^{rs}}(\vec{k},-\vec{k}).
	\label{malda2-1PI}
\end{align}
In momentum space~\1eq{con1PI-1zeta2gamma} reads
\begin{align}
	\langle\zeta(\vec{q}\,)\gamma^{\lambda_1}(\vec{k}\,)\gamma^{\lambda_2}(\vec{p}\,)\rangle&=\langle\zeta(\vec{q}\,)\zeta(-\vec{q}\,)\rangle\sum_{\lambda,\lambda'=\pm}\epsilon^\lambda_{ab}(\vec{k}\,)\epsilon^{\lambda'}_{mn}(-\vec{k})\langle\gamma^{\lambda'}(\vec{k}\,)\gamma^\lambda(-\vec{k}\,)\rangle\nonumber \\
	&\times\sum_{\lambda,\lambda'=\pm}\epsilon^\lambda_{cd}(\vec{p}\,)\epsilon^{\lambda'}_{rs}(\vec{p}\,)\langle\gamma^{\lambda'}(-\vec{p}\,)\gamma^\lambda(\vec{p}\,)\rangle\Gzt_{\zeta\gamma^{mn}\gamma^{rs}}(\vec{q},\vec{k},\vec{p}\,)\epsilon_{ab}^{\lambda_1}(\vec{k}\,)\epsilon_{cd}^{\lambda_2}(\vec{p}\,).
\end{align}
Substituting the result~\noeq{malda2-1PI} into the above equation, we obtain for the corresponding connected amplitude in the squeezed limit
\begin{align}
	\langle\zeta(\vec{q}\,)\gamma^{\lambda_1}(\vec{k}\,)\gamma^{\lambda_2}(\vec{p}\,)\rangle&\underset{\vec{q}\to0}{\sim}-n_t\langle\zeta(\vec{q}\,)\zeta(-\vec{q}\,)\rangle\sum_{\lambda,\lambda'=\pm}\epsilon^\lambda_{ab}(\vec{k}\,)\epsilon^{\lambda'}_{cd}(-\vec{k})\langle\gamma^{\lambda'}(\vec{k}\,)\gamma^\lambda(-\vec{k}\,)\rangle\epsilon_{ab}^{\lambda_1}(\vec{k}\,)\epsilon_{cd}^{\lambda_2}(\vec{-k}\,)\nonumber \\
	&=-4n_t r_1(\vec{k})\delta^{\lambda_1\lambda_2}\langle\zeta(\vec{q}\,)\zeta(-\vec{q}\,)\rangle.
\end{align} 

For completeness we report here the expression of the correlator $\langle \gamma^{\lambda'}\gamma^\lambda\rangle$ in terms of the graviton propagator:
\begin{align}
\langle\gamma^{\lambda'}(\vec{k}\,)\gamma^\lambda(-\vec{k}\,)\rangle =
\frac{1}{4} \epsilon^\lambda_{ij}(-\vec{k}\,) \epsilon^{\lambda'}_{mn}(\vec{k}\,) \Delta_{ij,mn}(\vec{k}) \, .
\end{align}
\subsection{Diffeomorphism ST identities}

To exhaust the three-point sector one needs to analyze two more correlators: $\langle\gamma\zeta\zeta\rangle$ and $\langle\gamma\gamma\gamma\rangle$, in which the graviton field is now soft. The corresponding consistency conditions have nothing to do with dilatation invariance, rather, as we will see, being a result of (the more general) diffeomorphism invariance. 

\subsubsection{One soft graviton and two scalars}

The amplitude we need to consider is in momentum space 
\begin{align}
	\langle\gamma^{\lambda_1}(\vec{q}\,)\zeta(\vec{k}\,)\zeta(\vec{p}\,)\rangle&=
	 \sum_{\lambda,\lambda'=\pm}\epsilon^{\lambda'}_{ab}(\vec{q}\,)\epsilon^\lambda_{mn}(-\vec{q}\,)\langle\gamma^{\lambda'}(\vec{q}\,)\gamma^\lambda(-\vec{q}\,)\rangle\langle\zeta(\vec{k}\,)\zeta(-\vec{k}\,)\rangle\nonumber \\
	&\times\langle\zeta(\vec{p}\,)\zeta(-\vec{p}\,)\rangle\Gzt_{\gamma^{mn}\zeta\zeta}(\vec{q},\vec{k},\vec{p}\,)\epsilon^{\lambda_1}_{ab}(\vec{q}\,)\nonumber \\
	&=2r_1(\vec{q}\,)\langle\zeta(\vec{k}\,)\zeta(-\vec{k}\,)\rangle\langle\zeta(\vec{p}\,)\zeta(-\vec{p}\,)\rangle\epsilon^{\lambda_1}_{mn}(\vec{q}\,)\Gzt_{\gamma^{mn}\zeta\zeta}(\vec{q},\vec{k},\vec{p}\,).
	\label{con1PI-1gamma2zeta}
\end{align}
Now observe that the 3-point function $\Gzt_{\gamma\zeta\zeta}$  admits the following form factor decomposition
\begin{align}
	\Gzt_{\gamma^{mn}\zeta\zeta}(\vec{q},\vec{k},\vec{p}\,)&=A\left(\delta^{mn}-\frac{q^m q^n}{q^2}\right)+B\frac{q^m q^n}{q^2}+C\frac{k^m k^n}{k^2}+D\left(\frac{k^m q^n}{q^2}+\frac{q^m k^n}{q^2}\right),
	\label{deco}
\end{align}
which means that the only form factor we are interested in is $C$ as
\begin{align}
	\epsilon^{\lambda_1}_{mn}(\vec{q}\,)\Gzt_{\gamma^{mn}\zeta\zeta}(\vec{q},\vec{k},\vec{p}\,)&=\epsilon^{\lambda_1}_{mn}(\vec{q}\,)\frac{k^m k^n}{k^2}C.
\end{align}
This holds as a consequence of the divergenceless condition $q^m \epsilon^\lambda_{mn}(q)=0$ and the traceless property $\epsilon^\lambda_{mm}(q)=0$ of the graviton polarizations.

This form factor can be determined at the lowest perturbative level by resorting to the ST identity satisfied by the 3-point function $\Gzt_{\gamma\zeta\zeta}$, which has been derived in~\1eq{3sf}. This clearly indicates that the sought for identity cannot be a consequence of dilatation invariance as we know that for such transformations $\Gzt_{c\gamma^*}$ vanishes, and therefore no information can be extracted for $\Gzt_{\gamma\zeta\zeta}$. 

However, as a result of diffeomorphism invariance, one has
\begin{align}
\left(\frac23i\delta_{mn}q^\ell-iq_n\delta_m^\ell-iq_m\delta_n^\ell\right)\Gzt_{\gamma^{mn}\zeta\zeta}(\vec{q},\vec{k},\vec{p}\,)&=\frac13iq^\ell\Gzt_{\zeta\zeta\zeta}(\vec{q},\vec{k},\vec{p}\,)+ik^\ell\Gzt_{\zeta\zeta}(\vec{p}\,)+ip^\ell\Gzt_{\zeta\zeta}(\vec{k}\,)
\end{align}
which, using the decomposition~\noeq{deco} leads to the two conditions (we write $\vec{p}=-\vec{q}-\vec{k}$)
\begin{align}
	&\frac43(A-B)+\frac23C-\frac23\frac{\vec{q}\cdot\vec{k}}{q^2}D=\frac13\Gzt_{\zeta\zeta\zeta}(\vec{q},\vec{k},-\vec{q}-\vec{k}\,)-\Gzt_{\zeta\zeta}(\vec{k}\,)\nonumber \\
	&-2\frac{\vec{q}\cdot\vec{k}}{k^2}C-2D=\Gzt_{\zeta\zeta}(\vec{q}+\vec{k}\,)-\Gzt_{\zeta\zeta}(\vec{k}\,).
	\label{ffacts}
\end{align}
Using the fact that the two-point function is a function of $k$ only, the rhs of the last equation above admits the expansion
\begin{align}
	\Gzt_{\zeta\zeta}(\vec{q}+\vec{k}\,)-\Gzt_{\zeta\zeta}(\vec{k}\,)&=\vec{q}\spr\vec{\partial}_k\Gzt_{\zeta\zeta}(\vec{k}\,)+{\cal O}(q^2)\nonumber\\
		&=\frac{\vec{q}\spr\vec{k}}{k^2}\left[k\partial_k \Gzt_{\zeta\zeta}(\vec{k}\,)\right]+{\cal O}(q^2).
\end{align}
implying that, in the squeezed limit $\vec{q}\to0$, the second equation in~\noeq{ffacts} will yield 
\begin{align}
	2D\underset{\vec{q}\to0}{\sim}-\frac{\vec{q}\cdot\vec{k}}{k^2}\left[2C+k\partial_k \Gzt_{\zeta\zeta}(\vec{k}\,)\right].
	\label{fres}
\end{align}
From the hypothesis of the attractor background as well as the adiabaticity condition~\cite{Berezhiani:2013ewa}, we know that $\widetilde\G^{(0)}$ is analytical in the squeezed limit. It then follows that the form factors $A$ and $B$ must be equal in this limit; then, substituting the result~\noeq{fres} into the first equation of~\noeq{ffacts}, will give to leading order
\begin{align}
	C&\underset{\vec{q}\to0}{\sim}\frac32\frac1{1+\cos^2(\vec{q},\vec{k}\,)}\left[\frac13\Gzt_{\zeta\zeta\zeta}(0,\vec{k},-\vec{k}\,)-\Gzt_{\zeta\zeta}(\vec{k}\,)-\frac13\cos^2(\vec{q},\vec{k}\,)k\partial_k \Gzt_{\zeta\zeta}(\vec{k}\,)\right].
\end{align}
where $(\vec{q},\vec{k}\,)$ represents the angle between the vectors $\vec{q}$ and $\vec{k}$.

Using finally the result~\noeq{res1}, we get
\begin{align}
	\epsilon^{\lambda_1}_{mn}(\vec{q}\,)\Gzt_{\gamma^{mn}\zeta\zeta}(\vec{q},\vec{k},\vec{p}\,)&\underset{\vec{q}\to0}{\sim}-{\epsilon^{\lambda_1}_{mn}(\vec{q}\,)k^m k^n}[\partial_{k^2}\Gzt_{\zeta\zeta}(\vec{k}\,)],
\end{align}
or, at the connected level\footnote{We use ${W}(t)_{J_\Phi J_\Phi}\widetilde{\Gamma}^{(0)}_{\Phi\Phi}=-1$, which implies \mbox{$\partial_{k^2}\langle\Phi\Phi\rangle=\langle\Phi\Phi\rangle[\partial_{k^2}\widetilde{\Gamma}^{(0)}_{\Phi\Phi}]\langle\Phi\Phi\rangle$}.}
\begin{align}
	\langle\gamma^{\lambda_1}(\vec{q}\,)\zeta(\vec{k}\,)\zeta(\vec{p}\,)\rangle&\underset{\vec{q}\to0}{\sim}-2r_1(\vec{q}\,){\epsilon^{\lambda_1}_{mn}(\vec{q})k^m k^n}[\partial_{k^2}\langle\zeta(\vec{k}\,)\zeta(-\vec{k}\,)\rangle].
\end{align}
%



\subsubsection{Three gravitons (one soft)}

The calculation of the three-graviton correlator amplitude proceeds in exactly the same way, even though its evaluation is a bit more complex from the algebraic point of view.
One starts from the amplitude
\begin{align}
	\langle\gamma^{\lambda_1}(\vec{q}\,)\gamma^{\lambda_2}(\vec{k}\,)\gamma^{\lambda_3}(\vec{p}\,)\rangle&=\sum_{\lambda,\lambda'=\pm}\epsilon^{\lambda'}_{ab}(\vec{q}\,)\epsilon^\lambda_{mn}(-\vec{q}\,)\langle\gamma^{\lambda'}(\vec{q}\,)\gamma^\lambda(-\vec{q}\,)\rangle\epsilon_{ab}^{\lambda_1}(\vec{q}\,)\nonumber \\
	&\times\sum_{\lambda,\lambda'=\pm}\epsilon^{\lambda'}_{cd}(\vec{k}\,)\epsilon^\lambda_{rs}(-\vec{q}\,)\langle\gamma^{\lambda'}(\vec{k}\,)\gamma^\lambda(-\vec{k}\,)\rangle\epsilon_{cd}^{\lambda_2}(\vec{k}\,)\nonumber \\
	&\times\sum_{\lambda,\lambda'=\pm}\epsilon^{\lambda'}_{ef}(\vec{p}\,)\epsilon^\lambda_{tu}(-\vec{p}\,)\langle\gamma^{\lambda'}(\vec{p}\,)\gamma^\lambda(-\vec{p}\,)\rangle\epsilon_{ef}^{\lambda_3}(\vec{p}\,)\Gzt_{\gamma^{mn}\gamma^{rs}\gamma^{tu}}(\vec{q},\vec{k},\vec{p}\,)\nonumber\\
	&=8r_1(\vec{q}\,)r_1(\vec{k}\,)r_1(\vec{p}\,)\epsilon^{\lambda_1}_{mn}(\vec{q}\,)\epsilon^{\lambda_2}_{rs}(\vec{k}\,)\epsilon^{\lambda_3}_{tu}(\vec{p}\,)\Gzt_{\gamma^{mn}\gamma^{rs}\gamma^{tu}}(\vec{q},\vec{k},\vec{p}\,).
	\label{con1PI-13gamma}
\end{align}
Notice that the amplitude $\epsilon^{\lambda_2}_{rs}\epsilon^{\lambda_3}_{tu}\Gzt_{\gamma^{mn}\gamma^{rs}\gamma^{tu}}$ admits exactly the form factor decomposition shown in~\1eq{deco}, so that one has\footnote{Obviously, in this case the form factors depend on the helicities $\lambda_2$ and $\lambda_3$} 
\begin{align}
	\epsilon^{\lambda_1}_{mn}(\vec{q}\,)\epsilon^{\lambda_2}_{rs}(\vec{k}\,)\epsilon^{\lambda_3}_{tu}(\vec{p}\,)\Gzt_{\gamma^{mn}\gamma^{rs}\gamma^{tu}}(\vec{q},\vec{k},\vec{p}\,)&=\epsilon^{\lambda_1}_{mn}(\vec{q}\,)\frac{k^m k^n}{k^2}C^{\lambda_2\lambda_3}
\end{align}

As in the previous case the form factor $C$ can be obtained by analyzing the ST identity constraining the function $\Gamma_{\gamma\gamma\gamma}$, which has been derived in~\1eq{1s2g}. As in the previous case one obtains the two conditions  
\begin{align}
	&\frac23C^{\lambda_2\lambda_3}-\frac23\frac{\vec{q}\cdot\vec{k}}{q^2}D^{\lambda_2\lambda_3}=\epsilon^{\lambda_2}_{rs}(\vec{k}\,)\epsilon^{\lambda_3}_{tu}(\vec{p}\,)\left[\frac13\Gzt_{\zeta\gamma^{rs}\gamma^{tu}}(\vec{q},\vec{k},-\vec{k}-\vec{q}\,)-\Gzt_{\gamma^{rs}\gamma^{tu}}(\vec{k}\,)\right]+\cdots\nonumber \\
	&-2\frac{\vec{q}\cdot\vec{k}}{k^2}C^{\lambda_2\lambda_3}-2D^{\lambda_2\lambda_3}=\epsilon^{\lambda_2}_{rs}(\vec{k}\,)\epsilon^{\lambda_3}_{tu}(\vec{p}\,)\frac{\vec{q}\cdot\vec{k}}{k^2}\left[k\partial_k \Gzt_{\gamma^{rs}\gamma^{tu}}(\vec{k}\,)\right]+\cdots,
\end{align}
where the dots stands for terms that vanish in the squeezed limit $\vec{q}\to0$. In this limit, using the result~\1eq{res2}, one then obtains at leading order 
\begin{align}
	\epsilon^{\lambda_1}_{mn}(\vec{q}\,)\epsilon^{\lambda_2}_{rs}(\vec{k}\,)\epsilon^{\lambda_3}_{tu}(\vec{p}\,)\Gzt_{\gamma^{mn}\gamma^{rs}\gamma^{tu}}(\vec{q},\vec{k},\vec{p}\,)&\underset{\vec{q}\to0}{\sim}-\epsilon^{\lambda_1}_{mn}(\vec{q}\,)\epsilon^{\lambda_2}_{rs}(\vec{k}\,)\epsilon^{\lambda_3}_{tu}(-\vec{k}\,)k^m k^n[\partial_{k^2}\Gzt_{\gamma^{rs}\gamma^{tu}}(\vec{k}\,)].
\end{align}

We now take the squeezed limit of~\1eq{con1PI-13gamma}, and substitute the result above  by taking into account that
\begin{align}
	2r_1(\vec{k}\,)\epsilon^{\lambda_2}_{rs}(\vec{k}\,)2r_1(\vec{k}\,)\epsilon^{\lambda_3}_{tu}(-\vec{k}\,)[\partial_{k^2}\Gzt_{\gamma^{rs}\gamma^{tu}}(\vec{k}\,)]&=\left[\partial_{k^2}\sum_{\lambda\lambda'=\pm}\epsilon^\lambda_{rs}(\vec{k}\,)\epsilon_{tu}^{\lambda'}(-\vec{k}\,)\langle\gamma^\lambda(\vec{k}\,)\gamma^{\lambda'}(-\vec{k}\,)\rangle\right]\nonumber \\
	&\times\epsilon^{\lambda_2}_{rs}(\vec{k}\,)
	\epsilon^{\lambda_3}_{tu}(-\vec{k}\,),
\end{align} 
so we finally get the result
\begin{align}
	\langle\gamma^{\lambda_1}(\vec{q}\,)\gamma^{\lambda_2}(\vec{k}\,)\gamma^{\lambda_3}(\vec{p}\,)\rangle&\underset{\vec{q}\to0}{\sim}-2r_1(\vec{q}\,)\epsilon^{\lambda_1}_{mn}(\vec{q}\,)\epsilon^{\lambda_2}_{rs}(\vec{k}\,)\epsilon^{\lambda_3}_{tu}(-\vec{k}\,)k^mk^n\nonumber \\
	&\times\left[\partial_{k^2}\sum_{\lambda\lambda'=\pm}\epsilon_\lambda^{rs}(\vec{k}\,)\epsilon^{tu}_{\lambda'}(-\vec{k}\,)\langle\gamma^\lambda(\vec{k}\,)\gamma^{\lambda'}(-\vec{k}\,)\rangle\right].
\end{align} 


\subsection{Special conformal transformations ST identities\label{SCTs}}

%

For the special conformal transformations, we adopt the same procedure used in the case of dilatations; therefore, we multiply the ST identity~\noeq{confgeneral} with~${\cal M}_{im}$ defined in~\1eq{sct}, integrate in $\mathrm{d}^3\vec{x}$, and, finally, Fourier transform the resulting expression.  

For the term proportional to the three-point function we obtain
\begin{align}
	\int\!\mathrm{d}^3\vec{x}\,{\cal M}_{im}(\vec{x}\,)\frac13\partial_{x^m}\Gzt_{\zeta\zeta\zeta}(\vec{x},\vec{y},\vec{z}\,)&=-2\int\!\mathrm{d}^3\vec{x}\,x_i\Gzt_{\zeta\zeta\zeta}(\vec{x},\vec{y},\vec{z}\,)\nonumber \\
	&=2i\int\!\frac{\mathrm{d}^3\vec{k}_1}{(2\pi)^3}\int\!\frac{\mathrm{d}^3\vec{k}_2}{(2\pi)^3}\mathrm{e}^{-i\vec{k_1}\cdot\vec{y}}\mathrm{e}^{-i\vec{k_2}\cdot\vec{z}}\left.\partial_{q^i}\Gzt_{\zeta\zeta\zeta}(\vec{q},\vec{k}_1,\vec{k}_2\,)\right\vert_{\vec{q}=0},
\end{align}
from which we see here that, as announced earlier, special conformal transformations determine (modulo a numerical factor) the next to leading order term in the squeezed limit expansion~\noeq{texp}.

The second term in~\noeq{confgeneral} gives
\begin{align}
	\int\!\mathrm{d}^3\vec{x}\,{\cal M}_{im}(\vec{x}\,)[-\partial_{x^m}\delta(\vec{x}-\vec{y})\Gzt_{\zeta\zeta}(\vec{x},\vec{z})]=[6y_i+{\cal M}_{im}(\vec{y}\,)\partial_{y^m}]\Gzt_{\zeta\zeta}(\vec{y},\vec{z}),
\end{align}
whereas the last term yields an identical result upon the replacement $\vec{y}\to\vec{z}$.
Taking this into account we find after Fourier transforming the two terms the results
%
\begin{align}
	6y_i\Gzt_{\zeta\zeta}(\vec{y},\vec{z}\,)+6z_i\Gzt_{\zeta\zeta}(\vec{z},\vec{y}\,)&=-6i\int\!\frac{\mathrm{d}^3\vec{k}_1}{(2\pi)^3}\int\!\frac{\mathrm{d}^3\vec{k}_2}{(2\pi)^3}\mathrm{e}^{-i\vec{k_1}\cdot\vec{y}}\mathrm{e}^{-i\vec{k_2}\cdot\vec{z}}\sum_{a=1}^2\partial_{k_a^i}\Gzt_{\zeta\zeta}(\vec{k}_1,\vec{k}_2),
\label{vanishingterm}
\end{align}
and
\begin{align}
	{\cal M}_{im}(\vec{y}\,)\partial_{y^m}\Gzt_{\zeta\zeta}(\vec{y},\vec{z})+{\cal M}_{im}(\vec{z}\,)\partial_{z^m}\Gzt_{\zeta\zeta}(\vec{z},\vec{y})&=i\int\!\frac{\mathrm{d}^3\vec{k}_1}{(2\pi)^3}\int\!\frac{\mathrm{d}^3\vec{k}_2}{(2\pi)^3}\mathrm{e}^{-i\vec{k_1}\cdot\vec{y}}\mathrm{e}^{-i\vec{k_2}\cdot\vec{z}}P_i\Gzt_{\zeta\zeta}(\vec{k}_1,\vec{k}_2).
\end{align}
where we have defined the operator
\begin{align}
	P_i&=\sum_{a=1}^2\left[6\partial_{k_a^i}-k_a^i\vec{\partial}_{k_a}^{\,2}+2\vec{k}_a\spr\vec{\partial}_{k_a}{\partial}_{k_a^i}\right].
	\label{sct-op}
\end{align}
Putting everything together one then obtains 
\begin{align}
	\left.\partial_{q^i}\Gzt_{\zeta\zeta\zeta}(\vec{q},\vec{k}_1,\vec{k}_2)\right\vert_{\vec{q}=0}=\frac12P_i\Gzt_{\zeta\zeta}(\vec{k}_1,\vec{k}_2)\Big\vert_{\vec{q}=0}-3\sum_{a=1}^2\partial_{k_a^i}\Gzt_{\zeta\zeta}(\vec{k}_1,\vec{k}_2)\Big\vert_{\vec{q}=0}.
	\label{res5}
\end{align}

To recover the result of~\cite{Creminelli:2012ed}, one needs to  consider the next to leading term in the squeezed limit  expansion of~\1eq{con1PI-3zeta.momspace}, or
\begin{align}
	\left.\partial_{q^i}\langle\zeta(\vec{k}_1)\zeta(-\vec{k}_1)\rangle\langle\zeta(\vec{k}_2)\zeta(-\vec{k}_2)\rangle\Gzt_{\zeta\zeta\zeta}(\vec{q},\vec{k}_1,\vec{k}_2)]\right\vert_{\vec{q}=0}.
\end{align}
It is then relatively easy to show that, using the leading order result~\noeq{res1}, that
\begin{align}
	\left.\partial_{q^i}\langle\zeta(\vec{k}_1)\zeta(-\vec{k}_1)\rangle\langle\zeta(\vec{k}_2)\zeta(-\vec{k}_2)\rangle\right\vert_{\vec{q}=0}\Gzt_{\zeta\zeta\zeta}(0,\vec{k},-\vec{k})=\,&
	6\,\partial_{k^i}\langle\zeta(\vec{k}\,)\zeta(-\vec{k}\,)\rangle+2k^j\Gzt_{\zeta\zeta}(\vec{k},-\vec{k}\,)\times\nonumber \\
	&\partial_{k^j}\langle\zeta(\vec{k}\,)\zeta(-\vec{k}\,)\rangle\partial_{k^i}\langle\zeta(\vec{k}\,)\zeta(-\vec{k}\,)\rangle,
	\label{part1}
\end{align}
whereas, using~\1eq{res5}, a lengthy (but otherwise straightforward) calculation shows that 
\begin{align}
	\langle\zeta(\vec{k}\,)\zeta(-\vec{k}\,)\rangle^2\left.\partial_{q^i}\Gzt_{\zeta\zeta\zeta}(0,\vec{k},-\vec{k})\right\vert_{\vec{q}=0}&=\frac12P_i\Gzt_{\zeta\zeta}(\vec{k}_1,\vec{k}_2)\Big\vert_{\vec{q}=0}-6\,\partial_{k^i}\langle\zeta(\vec{k}\,)\zeta(-\vec{k}\,)\rangle\nonumber \\
	&-2k^j\Gzt_{\zeta\zeta}(\vec{k},-\vec{k}\,)\partial_{k^j}\langle\zeta(\vec{k}\,)\zeta(-\vec{k}\,)\rangle\partial_{k^i}\langle\zeta(\vec{k}\,)\zeta(-\vec{k}\,)\rangle.
	\label{part2}
\end{align}
Summing up~\2eqs{part1}{part2} one obtains the final result~\cite{Creminelli:2012ed}
\begin{align}
	\left.\partial_{q^i}\langle\zeta(\vec{k}_1)\zeta(-\vec{k}_1)\rangle\langle\zeta(\vec{k}_2)\zeta(-\vec{k}_2)\rangle\Gzt_{\zeta\zeta\zeta}(\vec{q},\vec{k}_1,\vec{k}_2)]\right\vert_{\vec{q}=0}&=\frac12P_i\Gzt_{\zeta\zeta}(\vec{k}_1,\vec{k}_2)\Big\vert_{\vec{q}=0}.
\end{align}

It is evident that, due the fact that $\vec{k}_2=-\vec{k}_1$ the right-hand side of the expression above is zero, as is also zero the 1-PI result~\noeq{res5}. Thus for the three-point correlator, the next to leading term in the squeezed limit expansion vanishes, and the leading order corrections starts at ${\cal O}(q^2)$; and the latter will involve graviton corrections as at this order $(s\gamma)_{0}$ won't be zero anymore.

\section{Discussion and conclusions}\label{sec.mcc}

As we have shown in the last two sections, relations between 2- and 3-point correlators in the squeezed limit are nothing but a manifestation of ST identities. These latter identities in turn are derived from a single master consistency condition by taking one derivative with respect to the ghost field and then a certain number of derivatives with respect to fields with zero ghost number (not containing the $b$-field), setting afterwards all fields and external sources to zero. The generalization of the identities~\noeq{res1} and~\noeq{res5} for the scalar 1-PI $n$-point fixed-time functions is immediate, and one has
\begin{align}
	\Gzt_{\zeta\zeta_1\cdots\zeta_{n-1}}(0,\vec{k}_1,\dots,-\sum_{a=1}^{n-2}\vec{k}_a)&=[3(n-1)-\sum_{a=1}^{n-2}\vec{k}_a\spr\vec{\partial}_{k^a}]\Gzt_{_1\cdots\zeta_{n-1}}(\vec{k}_1,\dots,-\sum_{a=1}^{n-2}\vec{k}_a),\\
	\left.\partial_{q^i}\Gzt_{\zeta\zeta_1\cdots\zeta_{n-1}}(\vec{q},\vec{k}_1,\dots,\vec{k}_{n-1})\right\vert_{\vec{q}=0}&=\frac12P_i\Gzt_{\zeta_1,\dots\zeta_{n-1}}(\vec{k}_1,\dots,-\sum_{a=1}^{n-2}\vec{k}_a)\nonumber \\
	&-3\sum_{a=1}^{n-1}\partial_{k_a^i}\Gzt_{\zeta_1,\dots\zeta_{n-1}}(\vec{k}_1,\dots,-\sum_{a=1}^{n-2}\vec{k}_a).
	\label{res6}
\end{align}
with $P^i$ the trivial generalization of the operator~\noeq{sct-op}. What is much less obvious and possibly deserves further investigation, is the relation between these fixed time 1-PI functions and the corresponding $n$-point connected correlators, as $\Gzt(t)$ does not correspond to the effective action. 

Also, in renormalizable gauge theories the big advantage of the ST identity is that, for anomaly-free models, it survives quantization. Therefore the functional identity holding true for the tree-level action extends to the full vertex functional when radiative corrections are included. At higher orders in the loop expansion the BRST transformation gets renormalized and its effect on the master consistency condition is accounted for by the non-trivial corrections to the Green functions involving the insertion of an antifield and a ghost. When considering cosmological theories, however, the situation is much more involved. The ``in-in'' formalism requires a doubling of the field variables,  splitting them into left and right fields. This requires a reassessment of the analysis presented here, which avoided this splitting in view of the fact that for tree graphs at fixed times, left and right fields are the same.     

For the sake of completeness we work out in Appendix~\ref{app.gauge.indip.mast} the derivation of the identities stemming from the master consistency condition without replacing the antifield-ghost Green functions with their classical counterpart. These identities are bound to be the correct ones for any extension of the classical theory of gravity coupled to a scalar field when (even partial) radiative corrections are taken into account (see e.g.~\cite{NeferSenoguz:2008nn,Enqvist:2013eua}).

One should notice that the ST identity guarantees the physical unitarity of the theory, that is the cancellation of all ghost contributions. Therefore, any approximation to include quantum corrections (even within a fixed sector of the theory) needs to respect the ST identity in order to be consistent. If, on the other hand, the latter identity is broken, unphysical and potentially large numerical contributions may arise, as hinted at by the $1/\epsilon$ term in the unphysical part of the graviton propagator parameterized by $r_2$ [see~\1eq{grav.prop.params}]. 

The present formalism can be extended in a straightforward way to multi-field inflationary theories~\cite{Byrnes:2010em}. It turns out however that in general it is impossible to obtain consistency conditions arising from the ST identity and involving only correlators of the scalar and graviton metric components~(for a recent discussion, see, {\it e.g.}, \cite{Kenton:2015lxa}). This is because one can use the temporal gauge-fixing condition to eliminate only one inflaton scalar fluctuation, leaving in multi field scenarios the remaining contributions to the ST identity arising from gravitationally coupled scalars.


%
%
%


\appendix

\section*{Acknowledgments}

We thank the anonymous JCAP referee for pointing out an error in an earlier version of this paper, and for his critical remarks that have helped in expanding and elucidating many aspects of the framework proposed. One of us (A.Q.) would like to thank the warm hospitality of Prof. K.~Costello at Perimeter Institute, where an early version of this project was started.

\section{Quadratic Part of the ADM Action}\label{app.quadratic.part.adm}

We expand the ADM action in \1eq{ADMSi} up to second order in the fluctuations around the classical background.
One obtains, for the gravitational part
\begin{align}
	S_1^{(2)} = -\frac{1}{4}\int\!{\mathrm d}t\!\int\! {\mathrm d}^3 \vec{x} \,e^{\rho} &\bigg[\frac{1}{2} \partial^k \deltag^{ij} \partial_k \deltag_{ij} - \frac{1}{2} \partial^k \deltag_i^i \partial_k \deltag_j^j+ \partial^k \deltag_i^i \partial^j \deltag_{kj} - \partial^i \deltag^{kj} \partial_j \deltag_{ki}\nonumber \\
	&-2\delta{\cal N}( \partial^i \partial^j \deltag_{ij} - \partial^2 \deltag_i^i )\bigg],\nonumber\\
	S^{(2)}_2 =\int\!{\mathrm d}t\!\int\! {\mathrm d}^3 \vec{x} \, e^{\rho}&\bigg[-3e^{2 \rho}\dot{\rho}^2 \delta {\cal N}^{\,2} +\delta{\cal N} \left( \frac{3}{2} e^{2 \rho} \dot{\rho}^2 \deltag_i^i -2 \dot{\rho}\, \partial^i\,  {\cal N}_i + e^{2\rho}  \dot{\rho}\,  \dot\deltag_i^i \right) \nonumber \\
& + \frac{3}{8}  e^{2 \rho} \left(2 \deltag^{ij} \deltag_{ij} - \deltag_i^i \deltag_j^j\right) \dot{\rho}^2 
+ \frac12e^{2\rho}\left(2\deltag^{ij}\dot{\deltag_{ij}}-\deltag_i^i\dot{\deltag_j^j}\right)\dot{\rho} \nonumber \\
& -\frac{1}{4}    ( \partial_{[i} {\cal N}_{j]} \dot\deltag_{ij} - 2 \partial^i\, {\cal N}_i \dot\deltag_j^j )+ e^{-2\rho} \left ( \frac{1}{8} \partial^{[i}{\cal N}^{j]} \partial_{[i}{\cal N}_{j]} - \frac{1}{2} \partial^i\, {\cal N}_i \,\partial^j {\cal N}_j \right )\nonumber \\
& + e^{2\rho} \left ( \frac{1}{8} \dot\deltag^{ij} \dot\deltag_{ij} - \frac{1}{8} \dot\deltag_i^i \dot\deltag_j^j \right ) \bigg ],
\end{align}
where we have used the symmetrization convention $v_{[a}u_{b]} = v_a u_b + v_b u_a$. The inflaton part yields instead
\begin{align}
	S_3^{(2)} = -\frac{1}{2} \int\!{\mathrm d}t\!\int\! {\mathrm d}^3 \vec{x} \,  e^{3\rho}&\left\{-\bigg[\delta{\cal N}^{\,2}-\frac12\delta{\cal N}\deltag_i^i-\frac18( 2 \deltag^{ij} \deltag_{ij} - \deltag_i^i\deltag_j^j )
	\bigg]\dot{\varphi}^2\right.&\nonumber \\
	&+\left[2\delta{\cal N}-\deltag_i^i
	\right]\dot{\delta\phi}\,\dot{\varphi}\,-\dot{\delta\phi}^2-2e^{-2\rho}\dot{\varphi}\,\delta\phi\,\partial^i\,{\cal N}_i\nonumber \\
	&+V''(\varphi) \delta \phi^2+\left[ 2 V'(\varphi) \delta \phi + V(\varphi) \deltag_{ii} \right] \delta {\cal N}+V'(\varphi) \deltag_i^i \delta \phi \nonumber \\
	&-\left.\frac14 V(\varphi)(2 \deltag^{ij} \deltag_{ij} - \deltag_i^i\deltag_j^j) +e^{-2\rho}\partial^i \delta \phi \partial_i \delta \phi\right\}.
\end{align}
Due to the algebraic character of the equations of motion for the lapse and shift vector we can diagonalize the quadratic part by eliminating the mixing between ${\cal N}$, ${\cal N}^i$ and the other fields.
For that purpose we collect all the terms involving the fluctuations of the lapse function or the shift vector and set
\begin{align}
	S^{(2)}&=\widetilde{S}^{(2)}_1+\widetilde{S}_2^{(2)}+\widetilde{S}_3^{(2)}+S^{(2)}_\s{\delta{\cal N}\textrm{-}{\cal N}^i};& \widetilde{S}_i^{(2)}&=\left.S^{(2)}_i\right\vert_{\delta{\cal N}={\cal N}_i=0},
\end{align}
where
\begin{align}
	S^{(2)}_\s{\delta{\cal N}\textrm{-}{\cal N}^i}=\int\!{\mathrm d}t\!\int\! {\mathrm d}^3 \vec{x} \,&\left[a\,\delta{\cal N}^{\,2}+b\,\delta{\cal N}+c\,\delta{\cal N}\partial^i\,{\cal N}_i+d\,\partial^i\,{\cal N}_i+q^{ij}\partial_{[i}{\cal N}_{j]}\right.\nonumber \\
	&+\left.r\partial^{[i}{\cal N}^{j]}\partial_{[i}{\cal N}_{j]}+s\partial^i\, {\cal N}_i \,\partial^j {\cal N}_j\right], 
	\label{Nsector}
\end{align}
and we have defined the coefficients
\begin{align}
	a&=-\frac12e^{3\rho}(6\dot{\rho}^2-\dot{\varphi}^2);&
	b&=\frac12e^\rho( \partial^i \partial^j \deltag_{ij} - \partial^2 \deltag_i^i)+e^{3\rho}\dot{\rho}\dot{\deltag_i^i}-e^{3\rho}\dot{\varphi}\,\dot{\delta\phi}-e^{3\rho}V'(\varphi)\delta\phi;\nonumber \\
	c&=-2e^{\rho}\dot{\rho};&
	d&=\frac12e^{\rho}(\dot{\deltag_i^i}+2\dot{\varphi}\,\delta\phi);\nonumber \\
	q_{ij}&=-\frac14e^\rho\dot{\deltag_{ij}};&
	r&=\frac18e^{-\rho};\nonumber \\
	s&=-\frac12e^{-\rho}.
	\label{coeff1}
\end{align}
Then, after the change of variable
\begin{align}
	\delta{\cal N}'&=\delta{\cal N}+\frac1{2a}(b+c\,\partial^i\,{\cal N}_i),
\end{align}
the action~\1eq{Nsector} becomes
\begin{align}
	S^{(2)}_\s{\delta{\cal N}\textrm{-}{\cal N}_i}&=\int\!{\mathrm d}t\!\int\! {\mathrm d}^3 \vec{x} \,\left[-\frac{b^2}{4a}+a\,\delta{\cal N}'^{\,2}+(d'\delta^{ij}+2q^{ij})\,\partial_i\,{\cal N}_j+{\cal N}^i R_{ij} {\cal N}^j\right], 
\end{align}
where 
\begin{align}
	d'&=d-\frac{bc}{2a};& s'&=s-\frac{c^2}{4a};& R_{ij}&=-2r\delta_{ij}\partial^2 -(2r+s')\partial_i\partial_j.
\end{align}
Diagonalization is eventually achieved by making the change of variables
\begin{align}
	{\cal N}'_i&={\cal N}_i-\frac12R_{ij}^{-1}(\partial^jd'+2\partial_k q^{jk});&
	R_{ij}^{-1}&=-\frac1{2r\partial^2}\delta_{ij}+\frac1{4r+s'}\left(1+\frac{s'}{2r}\right)\frac{\partial_i\partial_j}{\partial^4},
\end{align}
yielding the final result
\begin{align}
	S^{(2)}_\s{\delta{\cal N}\textrm{-}{\cal N}_i}&=\int\!{\mathrm d}t\!\int\! {\mathrm d}^3 \vec{x} \,\left[-\frac{b^2}{4a}-\frac14(\partial^id'+2\partial_m q^{im})R^{-1}_{ij}(\partial^jd'+2\partial_n q^{jn})+\delta{\cal N}'a\,\delta{\cal N}'+{\cal N}'_i R^{ij} {\cal N}'_j\right].
	\label{Slapseshift}
\end{align}
Then, in the new primed variables the equations of motion of the lapse function and shift vector are simply given by
\begin{align}
	\delta{\cal N}'&=0 \qquad\Longrightarrow\qquad\delta{\cal N}=-\frac b{2a}-\frac c{2a}\partial^i{\cal N}_i;\nonumber \\
	{\cal N}'_i&=0 \qquad\Longrightarrow \qquad{\cal N}_i=\frac12R_{ij}^{-1}(\partial^jd'+2\partial_k q^{jk}),
\end{align}
which will leave the following contribution to the remaining terms in the action
\begin{align}
	\widetilde{S}^{(2)}_\s{\delta{\cal N}\textrm{-}{\cal N}_i}&=\int\!{\mathrm d}t\!\int\! {\mathrm d}^3 \vec{x} \,\left[-\frac{b^2}{4a}-\frac14(\partial^id'+2\partial_m q^{im})R^{-1}_{ij}(\partial^jd'+2\partial_n q^{jn})\right]\nonumber \\
	&=\int\!{\mathrm d}t\!\int\! {\mathrm d}^3 \vec{x} \,\left[\frac{d}{c^2}(ad-bc)+\frac{4a}{c^2}q^{im}\frac{\partial_i\partial_m}{\partial^2}\left(d'+\frac{\partial^j\partial^n}{\partial^2}q_{jn}\right)\right.\nonumber \\
	&\left.\hspace{2.5cm}+\frac{q_{im}}{2r}\frac{\partial^m\partial^n}{\partial^4}\left(\partial^i\partial^j-\delta^{ij}\partial^2\right)q_{jn}\right].
	\label{remainder}
\end{align}

We now choose a comoving type of gauge for the scalar sector by setting 
\begin{align}
	\delta\phi=0,
	\label{como_infl}
\end{align}

but leaving unspecified the gauge for the metric sector. Then we obtain
\begin{align}
	\widetilde{S}^{(2)}_\s{\delta{\cal N}\textrm{-}{\cal N}_i}&=\int\!{\mathrm d}t\!\int\! {\mathrm d}^3 \vec{x} \,\frac{e^{3\rho}}{16}\,\dot{\delta h}_{im}\left[\left(1+\frac{\dot{\varphi}^2}{2\dot\rho^2}\right)\left(\delta^{im}\delta^{jn}+\frac{\partial^i\partial^j\partial^m\partial^n}{\partial^4}\right)-4\delta^{ij}\frac{\partial^m\partial^n}{\partial^2}\right.\nonumber \\
	&\left.\hspace{4cm}+2\left(1-\frac{\dot{\varphi}^2}{2\dot\rho^2}\right)\delta^{jn}\frac{\partial^i\partial^m}{\partial^2}\right]\dot{\delta h}_{jn}\nonumber \\
	&-\int\!{\mathrm d}t\!\int\! {\mathrm d}^3 \vec{x} \,\frac{e^{\rho}}{8\dot\rho}\,\dot{\delta h}_{im}\frac1{\partial^2}\left(\partial^i\partial^m-\delta^{im}\partial^2\right)\left(\partial^j\partial^n-\delta^{jn}\partial^2\right)\delta h_{jn},
\end{align}
with the remaining terms of the action giving
\begin{align}
	\widetilde{S}_1^{(2)} &=\int\!{\mathrm d}t\!\int\! {\mathrm d}^3 \vec{x} \,\frac{e^{\rho}}8\delta h_{im} \left[\delta^{im}\left(2\partial^j\partial^n-\delta^{jn}\partial^2\right)-\delta^{ij}\left(2\partial^m\partial^n-\delta^{mn}\partial^2\right)\right]\delta h_{jn};\nonumber \\
	\widetilde{S}_3^{(2)} &=\int\!{\mathrm d}t\!\int\! {\mathrm d}^3 \vec{x} \,\frac{e^{3 \rho}}8 \left[4\dot\rho\left(2\delta h^{ij}\dot{\delta h_{ij}}-\delta h_i^i\dot{\delta h_j^j}\right)+2V\left(2\delta h^{ij}\delta h_{ij}-\delta h_i^i\delta h_j^j\right)\right.\nonumber \\
	&\left.+\dot{\delta h^{ij}}\dot{\delta h_{ij}}-\dot{\delta h_i^i}\dot{\delta h_j^j}\right].
\end{align}

If we were to choose the comoving gauge also for the metric sector we would set
\begin{align}
	\deltag_{ij}&=2\zeta\delta_{ij}+\gamma_{ij};&
	\partial^i\gamma_{ij}&=0;&\gamma_i^i&=0.
	\label{como}
\end{align}
This would yield the lapse function and shift vector solutions\footnote{A second gauge choice could have been
	$$\delta\phi=\widetilde{\varphi}(t,\vec x);\qquad \deltag_{ij}=\gamma_{ij};\qquad
	\partial_i\gamma_{ij}=0;\qquad\gamma_{ii}=0,$$
	in which case one immediately obtains 
	$$\delta N=\frac{\dot{\varphi}}{2\dot\rho}\widetilde{\varphi};\qquad {\cal N}_i=\partial_i\chi;\qquad
	\partial^2\chi=e^{2\rho}\frac{\dot{\varphi}^2}{2\dot\rho^2}\frac{\mathrm d}{{\mathrm d}t}\left[-\frac{\dot\rho}{\dot{\varphi}}\widetilde{\varphi}\right].
	$$}
\begin{align}
	\delta{\cal N}&=\frac{\dot\zeta}{\dot\rho};&
	{\cal N}_i&=\partial_i\psi;&
	\psi&=-\frac\zeta{\dot\rho}+\chi;&
	\partial^2\chi&=e^{2\rho}\frac{\dot{\varphi}^2\dot\zeta}{2\dot\rho^2}.
	\label{lapse.shift.vec.eom}
\end{align}
In addition, it is lengthy but straightforward to observe that
\begin{align}
	\widetilde{S}^{(2)}_\s{{\cal N},{\cal N}_i}
	&=\int\!{\mathrm d}t\!\int\! {\mathrm d}^3 \vec{x} \,e^{3\rho}\left[3\dot\zeta^2+\frac{\dot{\varphi}^2\dot\zeta^2}{2\dot\rho^2}-2e^{-2\rho}\frac{\dot\zeta}{\dot\rho}\partial^2\zeta\right],
\end{align} 
{\it i.e.}, as expected tensor fluctuations do not contribute to the scalar part of the action. Then one immediately obtains
for the quadratic parts of the action in the scalar fluctuation $\zeta$  $\left.\widetilde{S}^{(2)}\right\vert_{\zeta}$ and in the vector fluctuation $\left.\widetilde{S}^{(2)}\right\vert_{\gamma}$  the following results, up to irrelevant total derivative terms
\begin{align}
	\left.\widetilde{S}^{(2)}\right\vert_{\gamma}&=\frac{1}{8} \int\!{\mathrm d}t\!\int\! {\mathrm d}^3 \vec{x} \,  \left[e^{3\rho}\dot{\gamma}^{ij}\dot{\gamma}_{ij}-e^{\rho} \partial^k \gamma^{ij} \partial_k \gamma_{ij}\right],\nonumber \\
	\left.\widetilde{S}^{(2)}\right\vert_{\zeta}&=\frac12\int\!{\mathrm d}t\!\int\! {\mathrm d}^3 \vec{x} \,\frac{\dot{\varphi}^2}{\dot\rho^2}\left[e^{3\rho}\dot\zeta^2-e^\rho(\partial\zeta)^2\right].
\end{align}
They match those of~\cite{Maldacena:2002vr}.

\section{The metric propagator}\label{app.metric.prop}

\subsection{Bi-tensor coefficients of the metric two-point function}

The coefficients $\theta_\alpha$ in the expansion of the metric two-point function on the basis
of the bi-tensors in Eq.(\ref{bitensors}) are given by the following differential operators:
\begin{align}
	\theta_1&=-\frac{e^{3\rho}}{16}\left[\partial_t^2-5\dot\rho\partial_t-(4V+ e^{-2\rho}\partial^2)\right];\nonumber \\
	\theta_2&=\frac{e^{3\rho}}{16}\bigg(1-\frac{\dot{\varphi}^2}{2\dot\rho^2}\bigg)\partial_t^2+\frac{e^{3\rho}}{16}\bigg[-5\dot\rho-3\frac{\dot{\varphi}^2}{2\dot\rho}-\frac{\ddot{\varphi}\dot{\varphi}}{\dot\rho^2}+\frac{\dot{\varphi}^2\ddot\rho}{\dot\rho^3}-2\frac{e^{-2\rho}}{\dot\rho}\partial^2\bigg]\partial_t-\frac{e^{3\rho}}{4}V-\frac{e^{\rho}}{8}\partial^2;\nonumber \\
	\theta_3&\equiv\theta_4=\frac{e^{3\rho}}{16}\partial_t^2+\frac{e^{3\rho}}{16}3\dot\rho\partial_t-\frac{e^{\rho}}{16}\partial^2;\nonumber \\
	\theta_5&\equiv\theta_6=-\frac{e^{3\rho}}{16}\bigg(1-\frac{\dot{\varphi}^2}{2\dot\rho^2}\bigg)\partial_t^2-\frac{e^{3\rho}}{16}\bigg[3\dot\rho-3\frac{\dot{\varphi}^2}{2\dot\rho}-\frac{\ddot{\varphi}\dot{\varphi}}{\dot\rho^2}+\frac{\dot{\varphi}^2\ddot\rho}{\dot\rho^3}-\frac{e^{-2\rho}}{\dot\rho}\partial^2\bigg]\partial_t+\frac{e^{\rho}}{8}\partial^2;\nonumber \\
	\theta_7&=-\frac{e^{3\rho}}{16}\bigg(1+\frac{\dot{\varphi}^2}{2\dot\rho^2}\bigg)\partial_t^2-\frac{e^{3\rho}}{16}\bigg[3\dot\rho+3\frac{\dot{\varphi}^2}{2\dot\rho}+\frac{\ddot{\varphi}\dot{\varphi}}{\dot\rho^2}-\frac{\dot{\varphi}^2\ddot\rho}{\dot\rho^3}+2\frac{e^{-2\rho}}{\dot\rho}\partial^2\bigg]\partial_t.
	\label{thethetas}
\end{align}
In the second variation of the quadratic action with respect to the metric a new set operators $\Theta_\alpha$ appears; they read  
\begin{align}
	\Theta_\alpha=2c_\alpha(t)\partial_{t}^2+2\partial_{t}c_\alpha(t)\partial_{t}+[2a_\alpha(t)-\partial_{t}b_\alpha(t)+\partial_{t}^2c_\alpha(t)],
	\label{Theta1}
\end{align}
with the coefficients $a_\alpha$, $b_\alpha$ and $c_\alpha$ determined from the $\theta_\alpha$ operators in~\1eq{thethetas} after casting them in the form
\begin{align}
	\theta_\alpha=c_\alpha(t)\partial^2_t+b_\alpha(t)\partial_t+a_\alpha(t).
	\label{Theta2}
\end{align}

The $\Theta_a$ are given by
\begin{align}
\label{BigTheta.Eq}
& \Theta_1 = - \Theta_3 = - \Theta_4 =  -\frac{1}{8} e^{3 \rho} (
   \partial_t^2 + 3 \dot{\rho} \partial_t ) + \frac{1}{8}  e^{\rho}  \partial^2 ; \\
& \Theta_2 = - \Theta_5 = - \Theta_6 = \frac{ e^{3  \rho} }{8}\Big ( 1 - \frac{\dot{\phi}^2}{2 \dot{\rho}^2}   \Big ) \partial_t^2 +
 \frac{ e^{3  \rho} }{8} \Big ( 3 \dot{\rho} 
- \frac{3}{2} \frac{\dot{\phi}^2}{\dot{\rho}} 
- \frac{ \dot{\phi} \ddot{\phi}}{\dot{\rho}^2} 
+ \frac{\ddot{\rho} \dot{\phi}^2}{ \dot{\rho}^3} \Big ) \partial_t 
- \frac{e^\rho}{8} \Big ( 1 + \frac{\ddot{\rho}}{ \dot{\rho}^2} \Big )  \partial^2;  \\
& \Theta_7 = -\frac{ e^{3  \rho} }{8}\Big ( 1 + \frac{\dot{\phi}^2}{2 \dot{\rho}^2}   \Big ) \partial_t^2 +
 \frac{ e^{3  \rho} }{8} \Big (- 3 \dot{\rho} 
- \frac{3}{2} \frac{\dot{\phi}^2}{\dot{\rho}} 
- \frac{ \dot{\phi} \ddot{\phi}}{\dot{\rho}^2} 
+ \frac{\ddot{\rho} \dot{\phi}^2}{ \dot{\rho}^3} \Big ) \partial_t 
+ \frac{e^\rho}{8} \Big ( 1 - \frac{\ddot{\rho}}{ \dot{\rho}^2} \Big )  \partial^2 \, .
\end{align}

The operator $\Theta_1$ is proportional to the usual one entering into the field equation
for the physical graviton modes (see e.g. Eq.~(5.2.20) of~\cite{Weinberg:2008zzc} once one takes
into account the fact that the scale factor $a$ equals $e^\rho$ in our notations).

\subsection{Differential equations for the metric propagator}\label{app.diff.eq.metric.prop}

The inversion of the metric quadratic action in the $b$-$\delta g$ sector is relatively straightforward. 

One starts by passing to (three-)momentum space, $-i\partial\to {\vec p}$. We need then to solve the following matrix equation 
\begin{align}
	\left[
	\begin{array}{cc}
		\widetilde{\Gamma}_{mn}^{'(2)\, ij}({\vec p}\,) & B_a^{ij}({\vec p}\,) \\
		B_{mn}^c(-{\vec p}\,)& 0
	\end{array}
	\right]
	\left[
	\begin{array}{cc}
		\Delta_{rs}^{mn}({\vec p}\,) & C^{mn}_d({\vec p}\,) \\
		C^a_{rs}(-{\vec p}\,)& D^a_d(-{\vec p}\,)
	\end{array}
	\right]	=
	\left[
	\begin{array}{cc}
		\frac12{\cal O}^{(1)\, ij}_{rs} & 0 \\
		0 & \delta^b_d
	\end{array}
	\right]
\end{align}
where~\1eq{NL.term} implies $B_a^{ij}({\vec p}\,)=ip^i\delta^j_a-\frac i3p_a\delta^{ij}$. One has therefore the following four conditions to satisfy:
\begin{align}
	& \widetilde{\Gamma}_{mn}^{'(2)\, ij}({\vec p}\,)\Delta_{rs}^{mn}({\vec p}\,)+B_a^{ij}({\vec p}\,)C^a_{rs}(-{\vec p}\,)=\frac12{\cal O}^{(1)\, ij}_{rs}, \label{c1} \\
	& \widetilde{\Gamma}_{mn}^{'(2)\, ij}({\vec p}\,)C^{mn}_d({\vec p}\,)+B_a^{ij}({\vec p}\,) D^a_d(-{\vec p}\,)=0, \label{c2}\\
	& B_{mn}^c(-{\vec p}\,)\Delta_{rs}^{mn}({\vec p}\,)=0,\label{c3}\\
	& B_{mn}^c(-{\vec p}\,)C^{mn}_d({\vec p}\,)=\delta^b_d.\label{c4}
\end{align}

\1eq{c3} is the statement of the transversality of the metric propagator; writing
\begin{align}
	\Delta_{rs}^{mn}({\vec p}\,)=\sum_{\alpha=1}^7r_{\alpha}(p){\cal O}^{(\alpha)\, mn}_{rs},
\end{align}
it implies that in this gauge only the functions $r_1$ and $r_2$ need to be determined, whereas
\begin{align}
	r_3=r_4=-r_5=-r_6=-r_7=-r_1.
\end{align}

The solution of~\1eq{c4} is instead characterized by a single function $X$, and reads
\begin{align}
	C^{mn}_d({\vec p}\,)=X(p)\delta^{mn}p_d+\frac i{p^2}\left(\delta^m_dp^n+\delta^n_dp^m\right)-\frac i{2p^4}p^mp^np_d.
\end{align} 
\1eq{c2} can be then used to fix $X$. Writing 
\begin{align}
	D^a_d({\vec p}\,)=\Xi_1(p)\delta^a_d+\Xi_2(p)p^ap_d,	
\end{align}
one obtains
\begin{align}
	&-p^2\Xi_1=2\left(2\Theta_1+3\Theta_2+\Theta_6\right),\\
	&-ip^2\Xi_2=\left(2\Theta_3+\Theta_4+3\Theta_5+\Theta_7\right)X+\frac i{p^2}\left(-\Theta_1+\Theta_3+\Theta_4+\frac32\Theta_5+\frac32\Theta_7\right),\\
	&\left(2\Theta_1+\frac32\Theta_2+\frac23\Theta_3+\frac23\Theta_4+\Theta_5+\Theta_6+\frac13\Theta_7\right)X
	=0.
	\label{Xeq}
\end{align}
Now, on the one hand, as the differential operator appearing in~\1eq{Xeq} has a term that does not involve any derivative with respect to $t$, then a possible solution is provided by $X=0$. On the other hand,~\1eq{c1} gives consistent equations only for $X=0$, in which case it yields the two independent relations 
\begin{align}
	2\Theta_1 r_1&=\frac12;&
	(2\Theta_1 + 3 \Theta_2 + \Theta_6)s &= \frac12.
	\label{r1eq}
\end{align}
where $s=r_1+r_2$. 
Summarizing, the inversion of the metric two-point sector is achieved by setting
\begin{align}
	\Delta^{mn}_{rs}(p)&=r_1\left[{\cal O}^{(1)\,mn}_{rs}-{\cal O}^{(3)\,mn}_{rs}-{\cal O}^{(4)\,mn}_{rs}+{\cal O}^{(5)\,mn}_{rs}+{\cal O}^{(6)\,mn}_{rs}+{\cal O}^{(7)\,mn}_{rs}\right]+r_2{\cal O}^{(2)\,mn}_{rs}, \label{grav.prop} \\
	C^{mn}_d(p)&=\frac i{p^2}\left(\delta^m_dp^n+\delta^n_dp^m\right)-\frac i{2p^4}p^mp^np_d, \\
	D^a_d(p)&=-\frac2{p^2}\delta^a_d\left(2\Theta_1+3\Theta_2+\Theta_6\right)-\frac1{p^4}p^ap_d\left(-\Theta_1+\Theta_3+\Theta_4+\frac32\Theta_5+\frac32\Theta_7\right),
\end{align}
with $r_{1,2}$ determined by the differential equations above. 

\section{Identities for the connected generating functional}\label{app.funct}

The connected Green functions are generated by the functional $\cal{W}$, which is formally given by the 
Legendre transform of the vertex functional $\G$ with respect to the fields. $\G$ is a functional of the fields of the theory, which we
collectively denote by $\Phi$, and the external sources (e.g. the anti-fields, or other sources coupled to composite operators), that are denoted by $\Sigma$.
One has then
\begin{align}
{\cal W}[J_\Phi,\Sigma] = \Gamma[\Phi,\Sigma] + \int\!\d t\int\! \d^3 \vec{x}\, J_\Phi(t,\vec{x}\,) \Phi(t,\vec{x}\,)
\end{align}
where the source $J_\zeta$, coupled to the field $\zeta$, satisfies the conditions
\begin{align}
\label{src.j}
& J_\Phi(t,\vec{x}\,) = - \frac{\delta \G}{\delta \Phi(t,\vec{x}\,)} ,  & 
\Phi(t,\vec{x}\,) = \frac{\delta {\cal W}}{\delta J_\Phi(t,\vec{x}\,)} . 
\end{align}
On the other hand, for the external sources $\Sigma$ the following relation holds true
\begin{align}
\label{src.beta}
\frac{\delta \G}{\delta \Sigma(t,\vec{x}\,)} = \frac{\delta {\cal W}}{\delta \Sigma(t,\vec{x}\,)}. 
\end{align}
The temporal $b$-equation reads at the connected level
\begin{align}
-J_{b_0}(t,\vec{x}\,) = \frac{\delta {\cal W}}{\delta J_{\delta \phi}(t,\vec{x}\,)}.
\label{temp.beq.conn}
\end{align}
In the above equation we have denoted by $J_{b_0}$ the source coupled to $b_0$ and similarly for $J_{\delta \phi}$. Let us now take a derivative with respect to $J_{\delta \phi}$ of~\1eq{temp.beq.conn}. One then sees that the $\delta \phi$-propagator is zero (we set the external sources equal to zero after differentiation):
\begin{align}
\frac{\delta^2 {\cal W}}{\delta J_{\delta \phi}(t,\vec{x}\,) \delta J_{\delta \phi}(t,\vec{y}) } = 0 .
\label{deltaphi.prop}
\end{align}

In a similar way, by  taking a derivative of eq.(\ref{temp.beq.conn}) with respect to $J_{\delta h^{ij}}$ and then setting the sources to zero one has
\begin{align}
\frac{\delta^2 {\cal W}}{\delta J_{\delta h_{ij} }(t,\vec{x}\,) \delta J_{\delta \phi}(t,\vec{y}) } = 0 \, .
\label{deltaphi.graviton.prop}
\end{align}
Finally, by taking a derivative with respect to $J_b$ of eq.(\ref{temp.beq.conn}) we obtain
\begin{align}
\frac{\delta^2 {\cal W}}{\delta J_{b_0}(t,\vec{x}\,) \delta J_{\delta \phi}(t,\vec{y}) } = - \delta(t-t') \delta^{(3)}(\vec{x} - \vec{y})\, .
\label{deltaphi.b.prop}
\end{align}
Hence the only non-vanishing propagator involving $\delta \phi$ is the mixed $b$-$\delta \phi$ one. Moreover, by the temporal $b$-equation we see that there are no interaction vertices involving $b$ and therefore one can safely set $\delta \phi=0$ everywhere in the effective action. While this is expected in a comoving gauge, this is actually a proof that the $\delta \phi=0$ condition can indeed be consistently chosen.

The spatial $b$-equation reads at the connected level
\begin{align}
-J_{b^i} = \partial^k \frac{\delta {\cal W}}{\delta J_{\delta h_{ki}}(t,\vec{x}\,)} - \frac{1}{3} \partial_i \frac{\delta {\cal W}}{\delta J_{\delta h^k_k}(t,\vec{x}\,)} \, .
\label{spat.b.eq}
\end{align}
By taking a derivative with respect to $\delta J_{\delta h^{mn}}$ and then setting the sources to zero one gets the transversality condition for the metric propagator of~\1eq{c3}. The mixed $\delta \phi$-$\delta h$ propagator is zero, as we already know. Finally by taking a derivative of~\1eq{spat.b.eq} with respect to $J_{b^i}$ one gets back~\1eq{c2}.

\subsection*{Effective action}

The equations of motion for the lapse function and the shift vector are algebraic. It is therefore customary to
replace in the action $\delta {\cal N}$ and ${\cal N}^i$ with the solution of their equations of motion.
In the functional formalism this is equivalent to set the sources $J_{\delta {\cal N}}$ and 
$J_{{\cal N}^i}$ to zero in ${\cal W}$ and then take a Legendre transform back to the effective action
$\widetilde \G$:
\begin{align}
\widetilde \G[\delta \phi,\delta h, b, \bar c, c; \dots] = \left . {\cal W} \right |_{J_{\delta {\cal N}} = J_{{\cal N}^i} = 0} -
{\sum_{\Phi}}' \int\! \d t\int\!\d^3 \vec{x} \, J_{\Phi} (t,\vec{x}\,) \Phi(t,\vec{x}\,)
\end{align}
where the primed sum now runs over the fields with the exception of the lapse and the shift vector and the dots in the arguments of $\widetilde \G$ stand for the anti-fields and possibly other external sources coupled to composite operators.

By looking at \2eqs{src.j}{src.beta} it is easy to see that the effective action $\widetilde \G$ satisfies the following ST identity:
\begin{align}
{\cal S}(\widetilde \G) = \int\!\d t\int\! \d^3 \vec{x}\, \left[
\Gt_{\delta h^{ij*}}\Gt_{\delta h_{ij}} 
+ \Gt_{\phi^*} \Gt_{\phi}+ b^\mu \Gt_{\bar c^\mu} + 
\Gt_{c^{\mu *} } \Gt_{c_\mu } 
\right]  = 0.
\end{align}

\section{All order ST identity}\label{app.gauge.indip.mast}

For completeness, in this Appendix we collect the general form of both gauge independent as well as gauge dependent 
relations arising from the ST  identity. 

\subsection{Gauge independent identities}

As remarked in Sect.~\ref{sec.ST}, these relations are obtained by taking functional derivatives with respect to field combinations that do not involve the Nakanishi Lautrup multiplier $b$. Some of the following identities have been used in their classical approximation to recover Maldacena's consistency conditions in the squeezed limit (see Sect.~\ref{sec.maldacena}). 

We will give explicit expressions for all identities involving up to three insertions of $\delta \phi$ and $\delta h_{ij}$. To obtain them, we first take a derivative with respect to the ghost field $c$ and then take the derivative with respect to the field combination indicated (the notation $\widetilde \G_{\zeta_1 \dots \zeta_n \Phi^*_1 \dots \Phi^*_n}$ means that we evaluate the derivative of $\widetilde \G$ with respect to $\zeta_1 \dots \zeta_n, \Phi^*_1 \dots \Phi^*_n$  and then set the fields and external sources to zero).\\[10pt]
\noindent $(i)$ One derivative w.r.t $\delta h_{k l}$
\begin{align}
\label{master.1}
\int\left[\widetilde \G_{c^\rho \delta h^{ij*} \delta h_{k l}} \widetilde \G_{\delta h_{ij}}+
\widetilde \G_{c^\rho \delta h^{ij*}} \widetilde \G_{\delta h_{ij} \delta h_{k l}}+
\widetilde \G_{c^\rho \delta \phi^{*} \delta h_{k l}} \widetilde \G_{\delta \phi}+
\widetilde \G_{c^\rho \delta \phi^{*}} \widetilde \G_{\delta \phi \delta h_{k l}}\right] = 0.
\end{align}

\noindent $(ii)$ One derivative with respect to $\delta \phi$
\begin{align}
\label{master.2}
\int\left[\widetilde \G_{c^\rho \delta h^{ij*} \delta \phi} \widetilde \G_{\delta h_{ij}}+
\widetilde \G_{c^\rho \delta h^{ij*}} \widetilde \G_{\delta h_{ij} \delta \phi}+
\widetilde \G_{c^\rho \delta \phi^{*} \delta \phi} \widetilde \G_{\delta \phi}+
\widetilde \G_{c^\rho \delta \phi^{*}} \widetilde \G_{\delta \phi \delta \phi}\right] = 0.
\end{align}

\noindent $(iii)$ Two derivatives with respect to $\delta h_{k l}, \delta h_{m n}$
\begin{align}
\label{master.3}
&\int\left[ \widetilde \G_{c^\rho \delta h^{ij*} \delta h_{k l} \delta h_{m n}} \widetilde \G_{\delta h_{ij}}+
\widetilde \G_{c^\rho \delta h^{ij*} \delta h_{k l}} \widetilde \G_{\delta h_{ij} \delta h_{m n}}+
\widetilde \G_{c^\rho \delta h^{ij*} \delta h_{m n}} \widetilde \G_{\delta h_{ij} \delta h_{k l}}+
\widetilde \G_{c^\rho \delta h^{ij*}} \widetilde \G_{\delta h_{ij} \delta h_{k l} \delta h_{m n}}\right.
\nonumber \\
&\left.+ \widetilde \G_{c^\rho \delta \phi^{*} \delta h_{k l} \delta h_{m n}} \widetilde \G_{\delta \phi}+
\widetilde \G_{c^\rho \delta \phi^{*} \delta h_{k l}} \widetilde \G_{\delta \phi \delta h_{m n}}+
\widetilde \G_{c^\rho \delta \phi^{*} \delta h_{m n}} \widetilde \G_{\delta \phi \delta h_{k l}} +
\widetilde \G_{c^\rho \delta \phi^{*}} \widetilde \G_{\delta \phi \delta h_{k l} \delta h_{m n}}\right] = 0.
\end{align}

\noindent$(iv$ One derivative w.r.t $\delta h_{ij}$ and one with respect to $\delta \phi$
\begin{align}
\label{master.4}
&\int\left[ \widetilde \G_{c^\rho \delta h^{ij*} \delta \phi \delta h_{k l}} \widetilde \G_{\delta h_{ij}}+
\widetilde \G_{c^\rho \delta h^{ij*} \delta h_{k l}} \widetilde \G_{\delta \phi \delta h_{ij}}+
\widetilde \G_{c^\rho \delta h^{ij*} \delta \phi} \widetilde \G_{\delta h_{ij} \delta h_{k l}}+
\widetilde \G_{c^\rho \delta h^{ij*}} \widetilde \G_{\delta \phi \delta h_{ij} \delta h_{k l}}\right.
\nonumber \\
&\left.
+\widetilde \G_{c^\rho \delta \phi^{*} \delta \phi \delta h_{k l}} \widetilde \G_{\delta \phi}+
\widetilde \G_{c^\rho \delta \phi^{*} \delta h_{k l}} \widetilde \G_{\delta \phi \delta \phi}+
\widetilde \G_{c^\rho \delta \phi^{*} \delta \phi} \widetilde \G_{\delta \phi \delta h_{k l}} 
\widetilde \G_{c^\rho \delta \phi^{*}} \widetilde \G_{\delta \phi \delta \phi \delta h_{k l}}\right]= 0.
\end{align}

\noindent$(v)$ Two derivatives with respect to $\delta \phi$ [we set $\delta\phi_i=\delta\phi(z_i)$]
\begin{align}
\label{master.5}
&\int\left[
\widetilde \G_{c^\rho \delta h^{ij*} \delta \phi_2 \delta \phi_1} \widetilde \G_{\delta h_{ij}}+
\widetilde \G_{c^\rho \delta h^{ij*} \delta \phi_1} \widetilde \G_{\delta \phi_2 \delta h_{ij}}+
\widetilde \G_{c^\rho \delta h^{ij*} \delta \phi_2} \widetilde \G_{\delta h_{ij} \delta \phi_1}+
\widetilde \G_{c^\rho \delta h^{ij*}} \widetilde \G_{\delta h_{ij} \delta \phi_2 \delta \phi_1}\right. \nonumber \\
&\left.
+\widetilde \G_{c^\rho \delta \phi^{*} \delta \phi_2 \delta \phi_1} \widetilde \G_{\delta \phi}+
\widetilde \G_{c^\rho \delta \phi^{*} \delta \phi_1} \widetilde \G_{\delta \phi_2 \delta \phi}+
\widetilde \G_{c^\rho \delta \phi^{*} \delta \phi_2} \widetilde \G_{\delta \phi_1 \delta \phi} +
\widetilde \G_{c^\rho \delta \phi^{*}} \widetilde \G_{\delta \phi_2 \delta \phi_1 \delta \phi}\right]  = 0.
\end{align}

\subsection{Gauge-dependent identities}

We now take a derivative with respect to $c^\rho$ and w.r.t $b^\mu$ of~\1eq{eff.sti} and set the ghost field to zero.In order to be concrete we use the comoving gauge and the transverse gauge-fixing condition for the spatial part of the metric discussed in Sect.~\ref{sec.brst.gf}. One gets the following formula
\begin{align}
\int\Gt_{c^\rho  \bar c^\mu} = - \int\left[\left(- \delta_{j\mu} \partial_i + \frac{1}{3} \delta_{\mu s} \hat g_{ij} \partial_s \right)
\Gt_{c^\rho  h^*_{ij}} 
+ \delta_{\mu 0} \Gt_{\delta c^\rho \delta \phi^*} \right],
\label{ghost.sti}
\end{align}
where we have used the temporal and spatial $b$-equations in order to obtain the explicit form of $\Gt_{b^\mu \delta h_{ij}}$ and of $\Gt_{b^\mu \delta \phi}$, which depend on the specific gauge choice. We can further differentiate~\1eq{ghost.sti} with respect to $\delta h$ and $\delta \phi$, as being linear in the fluctuations, they cannot receive quantum corrections.

The relations found has a simple interpretation in this sector. Indeed, by taking the functional derivatives of both sides of Eq.(\ref{ghost.sti}) with respect to the metric or inflaton fluctuations, one sees that the Green function with one anti-ghost and one ghost and any number of $\delta h$ and $\delta \phi$ fields is determined in terms of the corresponding insertions on the rhs of~\1eq{ghost.sti} with the relevant anti-fields.
%

\end{document}